\documentclass[aps,prb,reprint, superscriptaddress]{revtex4-2}
\usepackage{graphicx}
\usepackage{dcolumn}
\usepackage{bm}
\usepackage{amsmath}
\usepackage{xcolor}
\allowdisplaybreaks

\begin{document}


\title{Modal decomposition of localized plasmon on gold nanoparticles}

\author{Gangcheng Yuan}
\email{gangcheng.yuan1@monash.edu}
\affiliation{%
 ARC Centre of Excellence in Exciton Science\\
 School of Chemistry, Monash University, Clayton, Victoria, 3800, Australia
}%

\author{Jared H. Cole}
\email{jared.cole@rmit.edu.au }
\affiliation{
 ARC Centre of Excellence in Exciton Science\\
 School of Science, RMIT University, Melbourne, 3001, Australia
}%
\author{Alison M. Funston}%
 \email{alison.funston@monash.edu}
\affiliation{%
 ARC Centre of Excellence in Exciton Science\\
 School of Chemistry, Monash University, Clayton, Victoria, 3800, Australia
}%

\date{\today}

\begin{abstract}
Localized surface plasmons (LSPs) are collective oscillations of free electrons in metal nanoparticles that confine electromagnetic waves into subwavelength regions, making them an ideal platform for light-matter coupling. To design and understand plasmonic structures, numerical computations of Maxwell's equations are commonly used. However, obtaining physical insight from these numerical solutions can be challenging, especially for complex-shaped nanoparticles.  To circumvent this, we introduce mode decomposition strategies within the boundary element method (BEM). By employing singular value decomposition (SVD) and quasi-normal mode (QNM) decomposition, we break down optical responses into elementary modes.  QNMs offer deeper insights into frequency and damping, while SVD modes allow for more accurate spectral reconstruction with fast computation. These techniques provide a deeper understanding of LSPs and facilitates the design of metal nanoparticles for efficient light-matter interaction.

\end{abstract}

\maketitle


\section{\label{sec:level1}Introduction}
Localized surface plasmons (LSPs) are collective oscillations of free electrons confined in a small region, typically found in metal nanoparticles with sizes smaller than the wavelength of light \cite{maier2007plasmonics,Amendola_2017}. LSPs can confine electromagnetic waves to regions below the diffraction limit, and their resonances are highly sensitive to both the nanoparticle's shape and the surrounding environment. These unique properties make LSPs highly versatile in applications such as sensing \cite{mayer_localized_2011,saha_gold_2012,stewart_nanostructured_2008,novo_direct_2008}, photocatalysis \cite{cortes_challenges_2020,zhou_light-driven_2020, christopher_visible-light-enhanced_2011,elias_elucidating_2022}, and the manipulation of light and its coupling with matter \cite{yu_flat_2014,schuller_plasmonics_2010,jiang_active_2018,chikkaraddy_single-molecule_2016,torma_strong_2014,HapuarachchiCampaioliCole+2022+4919+4927}.

The properties of LSPs can be characterized by these so-called resonance modes, however, their definition remains somewhat ambiguous. For simple geometries like spheres and cylinders, exact analytical solutions derived from Mie theory provide a clear determination of resonance modes \cite{bohren_absorption_1998}. The identification of resonance modes becomes cumbersome in the case of optical systems with arbitrary shapes where only numerical computation of Maxwell equations are feasible. Various numerical methods exist for this purpose \cite{myroshnychenko_modelling_2008,gallinet_numerical_2015}, such as the finite-difference time-domain method (FDTD) \cite{KY1138693,hao_plasmon_2007,taflove2005computational}, the finite element method (FEM) \cite{coggon_electromagnetic_1971,jin2015finite,hoffmann_comparison_2009}, discretized dipole approximation (DDA) \cite{purcell1973scattering,draine_discrete-dipole_1994}, and the boundary element method (BEM) \cite{PhysRevLett.80.5180,garcia_de_abajo_retarded_2002}. Typically, these methods involve exciting an optical system with a probe field and calculating the resulting optical response. Resonance modes are often identified by locating peaks in the extinction and scattering spectra, representing the maximum optical response. When a few resonance modes coexist, they can be separated from one another through empirical techniques such as global fitting with a set of Lorentzian functions \cite{C4CS00131A,B604856K} and non-negative decomposition \cite{nicoletti_three-dimensional_2013,pauca_nonnegative_2006,barrow_electron_2016}, which are commonly employed in the modal analysis of experimental optical spectra. However, the task of mode decomposition becomes challenging when multiple modes strongly overlap and interfere with each other, particularly in the case of large nanoparticles with complex geometries \cite{lukyanchuk_fano_2010,halas_plasmons_2011,hicks_controlling_2005}. Additionally, the identification of resonance modes heavily relies on the choice of either the probe field in simulations or the excitation conditions in experimental setups, leading to situations where not all relevant modes are discovered \cite{chu_probing_2009,gomez_dark_2013,bitton_vacuum_2020}.

Such an ambiguity of plasmon resonance modes can be effectively addressed by considering the natural resonance modes.  Natural resonance modes can sustain themselves in a optical system  in absence of any external excitation sources, and therefore they are intrinsic and solely depend on the system's characteristics \cite{baum_singularity_2005, StrattonJuliusAdams1941Et,bohren_absorption_1998}. The broadband optical response are dominated by these natural modes that can be excited by an external excitation. The application of natural modes traces back to the singularity expansion method, which was originally developed decades ago for calculating the electromagnetic response of antennas and scatterers \cite{1140398,1140603,baum_singularity_2005}. The singularity, or pole, is typically located at a complex frequency, with the imaginary part characterizing the damping of the corresponding mode. In recent years, due to the advancements in plasmonics, the increasing demand for novel nanophotonic cavities, and the development of effective algorithms for nonlinear eigenvalue problems, this concept has experienced a resurgence under the equivalent term "quasi-normal modes" (QNMs) \cite{sauvan_theory_2013,zheng_implementation_2014,kristensen_modeling_2020,kristensen_modes_2014,lalanne_light_2018}.  QNMs have been applied in estimating the Purcell effect and quality factor of nanoresonators \cite{sauvan_theory_2013,zheng_interacting_2013}, quantizing the cavity field \cite{franke_quantization_2019}, and comprehending the nonlinear response of plasmonic nanostructures \cite{bernasconi_mode_2016,gigli_quasinormal-mode_2020}.  In principle, an exact analysis of these resonance modes requires knowledge of the infinite-dimensional spectrum and eigenvectors of the relevant differential or integral operator governing the electromagnetic response \cite{hanson_operator_nodate,rahola_eigenvalues_2000,budko_spectrum_2006,budko_classification_2006}. To simplify this analysis, we can transform it into a more treatable finite-dimensional eigenvalue problem by utilizing a proper discretization method. The framework of QNMs has been implemented in various numerical methods in the frequency domain including FEM \cite{bai_efficient_2013},  BEM \cite{makitalo_modes_2014,unger_novel_2018,powell_resonant_2014}, as well as the time-domain method like FDTD \cite{kristensen_generalized_2012,kristensen_modes_2014}. Unlike the simple eigenvalue problems commonly found in quantum mechanics textbooks, the scattering or impedance matrix for general optical systems is usually non-Hermitian and nonlinear. For non-Hermitian open systems, the introduction of bi-orthogonality between left and right eigenvectors becomes necessary. The nonlinearity mainly arises from the dispersive permittivity and permeability of the materials \cite{sauvan_theory_2013,lalanne_light_2018,zolla_photonics_2018}, but it can also arise when the integral formalism \cite{makitalo_modes_2014,unger_novel_2018,powell_resonant_2014,lasson_three-dimensional_2013} is used with specific built-in boundary conditions, leading to nonlinear terms of frequency in the matrix elements. The presence of nonlinearities not only increases the difficulty of finding natural modes or eigenmodes but also raises concerns about the accuracy and reliability of the modal decomposition based on these modes.

Fortunately, for an object much smaller than the wavelength, the challenges posed by nonlinearity can be largely overcome when employing the quasi-static approximation. In this case,  Maxwell's equations can be simplified to an electrostatic Poisson equation \cite{stockman_localization_2001,mayergoyz_electrostatic_2005,ouyang_surface_1989,davis_designing_2009}. While an integral equation still arises from the Poisson equation, it can be discretized and then transformed into an equivalent linear eigenvalue problem where the frequency dependent part is isolated from the matrix and treated as a scalar number \cite{mayergoyz_electrostatic_2005,ouyang_surface_1989}. The natural resonance modes correspond to the eigen solutions that are frequency- and scale-invariant \cite{boudarham_modal_2012}. These quasistatic eigenmodes constitute a complete basis set, and the broadband optical response to the external excitation can be projected onto these eigenmodes \cite{boudarham_modal_2012,mayergoyz_electrostatic_2005,ouyang_surface_1989}. Electrostatic eigenmodes have proven effective in the modal analysis and decomposition of electron energy loss spectroscopy (EELS) \cite{RevModPhys.82.209,horl_tomography_2013,boudarham_modal_2012,hauer_tomographic_2023} and the explanation of  resonant mode interaction in plasmonics \cite{nordlander_plasmon_2004,lukyanchuk_fano_2010,davis_simple_2010}. When the system size increases to the scale of the wavelength and the impact of retardation effects becomes more pronounced, accurate computations require the consideration of the full set of Maxwell's equations, leading to a nonlinear eigenvalue problem. With a few exceptions, the completeness of the corresponding natural modes becomes a subject of doubt \cite{leung_completeness_1994,ching_quasinormal-mode_1998,lobanov_resonant-state_2019,mansuripur_leaky_2017}. While not as straightforward as in the electrostatic case, recent studies have managed to achieve an approximate decomposition of the optical response into a set of QNMs for full wave simulations \cite{sauvan_theory_2013,lalanne_light_2018,alpeggiani_quasinormal-mode_2017}. In essence, the original nonlinear matrix can be linearized through the introduction of auxiliary variables, resulting in a larger matrix where the elements are either frequency-independent or linearly dependent \cite{guttel_nonlinear_2017,yan_rigorous_2018,truong_continuous_2020,raman_photonic_2010}. If this subsequent matrix is diagonalizable, it can be decomposed into QNMs and additional numerical modes, forming a complete basis set. Beyond being a mere mathematical manipulation, however, providing a physical interpretation of this decomposition involving additional numerical modes can still pose significant challenges. It is important to note that such a decomposition may not be unique due to the non-orthogonality of QNMs and additional numerical modes \cite{gras_nonuniqueness_2020}.

Instead of seeking the complex natural frequency and its corresponding mode from a nonlinear matrix, it is possible to directly solve for a set of eigenvalues and eigenmodes (also known as characteristic modes)\cite{harrington_theory_1971,garbacz_generalized_1971,bergman_theory_1980,markel_antisymmetrical_1995} from the scattering or impedance matrix at each specific frequency. This approach involves solving a linear eigenvalue problem, which allows for straightforward decomposition of the scattering matrix using the eigenmodes. However, this modal analysis often needs to be performed for various frequencies, and both the eigenvalues and eigenmodes are functions of frequency \cite{makitalo_modes_2014,suryadharma_quantifying_2019}. It is worth noting that the modes associated with the smallest eigenvalues might have the most significant impact.  When the frequency equals to the natural frequency, the smallest eigenvalue becomes zero and the corresponding eigenmodes are the natural modes.

The potential-based boundary element method is widely used in nanoparticle plasmonics \cite{garcia_de_abajo_retarded_2002,RevModPhys.82.209,nelayah_mapping_2007,polman_electron-beam_2019,hohenester_mnpbem_2012,waxenegger_plasmonics_2015}. One of the key advantages of the boundary element method lies in its assumption of homogeneous dielectric functions for both metal nanoparticles and their surrounding environment. This simplification allows for the discretization of the boundaries only, significantly reducing the computational burden compared to the finite volume method. By requiring a much smaller number of elements, the BEM approach conserves computing resources while still providing accurate solutions for optical properties and interactions of nanoparticles. However, there have been limited investigations into mode decomposition based on this method \cite{alpeggiani_visible_2016,kristensen_modeling_2020}. In this study, we first discuss the properties of the BEM matrix, and the relation between left and right eigen vectors. Then we evaluate two different mode decomposition methods: singular value decomposition and QNMs decomposition. The singular value modes and QNMs are used as a Galerkin basis set to reduce the dimension of the BEM matrix. We give detailed explanations and provide examples of both decomposition methods, and compare their outcomes to Mie theory results. Furthermore, we establish connections between these decomposition strategies, highlighting both their similarities and differences. The overview of the mode decomposition methods is given by Table~\ref{tab:summary}.

\begin{table*}
\caption{\label{tab:summary}The overview of decomposition methods}
\centering
    \begin{ruledtabular}
        \begin{tabular}{llll}
            \multicolumn{1}{l}{Mode definition}&\multicolumn{1}{l}{Mode decomposition}&Computation cost& Accuracy\\\hline
            {Natural resonance modes, QNMs} & Direct decompositon Eq.~(\ref{eq:simpleexpansion}) &  low & low \\
            &calculation at a few resonance &&\\\cline{2-4}
            { $M(\omega^*)V(\omega^*)=0$}& Galerkin approximation Eq.~(B7) &  Medium & Medium \\
            &calculation at each frequency&&\\
            &the matrix dimension is reduced&&\\\hline
            {Characteristic modes} & SVD or EVD at each frequency & high & high \\
             &connect $V(\omega)$ by mode tracking algorithm &&\\\cline{2-4}
            {$M(\omega) = U(\omega)\Sigma(\omega)V(\omega)^{\dag}$ (SVD, EVD)} & Galerkin approximation Eq.~(\ref{svd1}) & low & high \\
               & SVD or EVD at a fixed frequency &  &\\
               & the matrix dimension is reduced &  &\\
        \end{tabular}
\end{ruledtabular}
\end{table*}

\section{\label{sec:level2}Decomposition method}
\subsection{BEM equations}
\begin{figure}[b]
\includegraphics{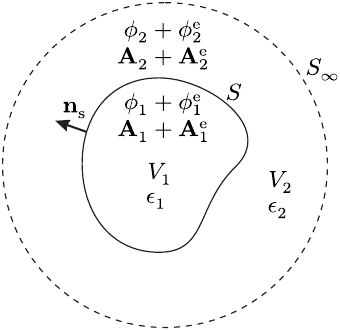}
\caption{\label{fig:BEM} Schematic of the boundary element method (BEM). A boundary surface denoted as $S$, with a normal vector $\mathbf{n}_s$, separates an internal region $V_1$ with a dielectric function $\epsilon_1$ from an outside background region $V_2$ with a dielectric function $\epsilon_2$. The boundary at infinity is represented as $S_{\infty}$. The electromagnetic boundary conditions establish the connection between the induced scalar and vector potentials, denoted as $\phi_{1,2}$ and $\mathbf{A}_{1,2}$, respectively, in the two regions, which are induced by the external scalar and vector potentials, $\phi^e_{1,2}$ and $\mathbf{A}_{1,2}^e$.}
\end{figure}

Before discussing modal decomposition, a brief introduction to the potential-based boundary element method (BEM) \cite{garcia_de_abajo_retarded_2002,RevModPhys.82.209} is presented. Additional details are available in Appendix~\ref{app:BEM} and the work by~\citet{garcia_de_abajo_retarded_2002}. This method has been implemented in the open-source MATLAB toolbox MNPBEM \cite{hohenester_mnpbem_2012,waxenegger_plasmonics_2015} and widely used for solving Maxwell's equations, particularly for addressing the plasmonic behavior of metallic nanoparticles with arbitrary shapes.  As shown in Fig.~\ref{fig:BEM}, BEM assumes that the dielectric function $\epsilon({\bf r},\omega)$ and magnetic permeability $\mu({\bf r}, \omega)$ are frequency-dependent, local, and isotropic. These properties remain uniform within distinct regions 1 and 2 but undergo abrupt changes at their boundaries. Electric and magnetic fields are expressed using scalar and vector potentials. $\phi_{1,2}(\bf r)$ and $\mathbf{A}_{1,2}(\bf r)$ can be mimicked by the surface charge $\sigma_{1,2}(s)$ and the surface current $\bf h_{1,2}(s)$ at the two sides of the boundary through the green function $G(r,s)$

\begin{equation}
    \phi({\bf r}) = G_{1,2}\sigma_{1,2} = \int_{S}d{\bf s}G_{1,2}(|{\bf r - s|})\sigma_{1,2}({\bf s}),
\end{equation}
and 
\begin{equation}
    {\bf A(r)} = G_{1,2}{\bf h}_{1,2} =\int_{S}d{\bf s}G_{1,2}(|{\bf r - s}|){\bf h}_{1,2}({\bf s}).
\end{equation}
Electromagnetic boundary conditions connect induced scalar and vector potentials in two regions, together with external potentials ($\phi^e_{1,2}$ and $\mathbf{A}_{1,2}^e$). This gives rise to the following equations for induced scalar and vector potentials ($\phi_{2}(\bf s)$ and $\mathbf{A}_{2}(\bf s)$) in the region 2 but infinitesimally close to the boundary $S$:
\begin{eqnarray}\label{foureqs}
\Gamma \phi_2 - ik(L_1-L_2){\bf n }_s\cdot \mathbf{A}_2 = {D}^{e*},\nonumber\\
\Delta \mathbf{A}_2 - ik(L_1-L_2){\bf n }_s \phi_2 = \vec{ \alpha}^{*}.
\end{eqnarray}
The definitions of the operators $\Gamma$, $\Delta$ and $L_{1,2}$ in Eqs.~(\ref{foureqs}) are given in Appendix~\ref{app:BEM}. $k$ is the wavenumber in the vacuum and ${\bf n }_s$ denotes the surface normal vector of the boundary at s, directed from 1 to 2.  ${D}^{e*}$ and $\vec{ \alpha}^{*}$ represent the external excitation source. For simplicity, all of these symbols can be understood as scalars, vectors and matrices after discretizing the boundary $S$ into $N$ pieces, and thus Eqs.~(\ref{foureqs}) can be approximated by a set of $4N$ linear equations with $4N$ variables.

After eliminating $\phi^e_{2}$, $\mathbf{A}_{2}(s)$ can be solved through a set of $3N$ linear equations:
\begin{eqnarray}\label{current_equations}
[\Delta + k^2(L_1-L_2){\bf n }_s \Gamma ^{-1} {\bf n }_s\cdot(L_1-L_2)]\mathbf{A}_{2}\nonumber\\
= \vec{ \alpha}^{*} + ik(L_1-L_2){\bf n }_s\Gamma ^{-1}{D}^{e*}.
\end{eqnarray}
Subsequently, we determine $\phi_2$ from
\begin{eqnarray}
\Gamma \phi_2 = D^{e*} + ik(L_1-L_2){\bf n}_s\cdot \mathbf{A}_{2}.
\end{eqnarray}
The potentials $\phi_1(s)$ and $\mathbf{A}_{1}(s)$ at the boundary from the inside can be computed either based on the outside potentials or by following a similar procedure. These potentials can be employed for the computation of the electric and magnetic fields. As described, the BEM approach transforms Maxwell's equations into a set of linear equations whose dimension is of the order of the boundary element number. This enables rapid computation of the optical response of metal nanoparticles. Additionally, the coefficient matrix of these linear equations can be utilized for mode definition and mode decomposition in the following parts.

\subsection{Modal Decomposition }
In this section, we introduce two modal decomposition approaches.  We express the surface current and vector potentials as series of modes, enabling the construction of electromagnetic fields and scattering spectra for gold nanoparticles subjected to external fields. We rewrite Eq. (\ref{current_equations}), which depends on frequency $\omega$, in the form as
\begin{eqnarray}\label{CurrentMatrix}
M(\omega)x(\omega)=y(\omega),
\end{eqnarray}
where
\begin{eqnarray}\label{Mdefiniion}
M(\omega)=\Delta I_3 + k^2(L_1-L_2){\bf n }_s \Gamma ^{-1} {\bf n }_s\cdot(L_1-L_2).
\end{eqnarray}
$I_3$ is the $3\times3 $ identity matrix;  $M(\omega)$ is a $3N\times 3N$ matrix function of $\omega$. The $3N$-dimensional vector $x(\omega)$ is  given by
\begin{eqnarray}\nonumber
x(\omega)=\begin{bmatrix}
  {A}_{2x}\\
  {A}_{2y}\\
  {A}_{2z}  
\end{bmatrix},
\end{eqnarray}
where the $N$-dimensional vectors $A_{2x,y,z}$ are the three components of $\mathbf{A}_{2}$ at the $N$ locations on the boundary. Similarly, the $3N$-dimensional external excitation source is given by 
\begin{eqnarray}\nonumber
y(\omega)=\begin{bmatrix}
  { \alpha}^{*}_{x} + ik(L_1-L_2){\bf n }_{sx}\Gamma^{-1}{D}^{e*}\\
  \alpha^{*}_{y} + ik(L_1-L_2){\bf n }_{sy}\Gamma^{-1}{D}^{e*}\\
  \alpha^{*}_{z} + ik(L_1-L_2){\bf n }_{sz}\Gamma ^{-1}{D}^{e*} 
\end{bmatrix}.
\end{eqnarray}

\subsubsection{Singular Value Decomposition }
In the first approach, we employ a straightforward singular value decomposition of $M(\omega)$, given by
\begin{eqnarray}\label{svd}
M(\omega) = U(\omega)\Sigma(\omega)V(\omega)^{\dag},
\end{eqnarray}
where the columns $u_{j=1,\dots,3N}(\omega)$ of the $3N \times 3N$ complex unitary matrix $U(\omega)$ are referred to as the left singular vectors. The columns $v_{j=1,\dots,3N}(\omega)$ of the $3N \times 3N$ complex unitary matrix $V(\omega)$ are known as the right singular vectors. Here, $V(\omega)^{\dag}$ represents the conjugate transpose of $V(\omega)$. Additionally, $\Sigma(\omega)$ is a $3N \times 3N$ diagonal matrix with non-negative real diagonal entries $\sigma_j^d(\omega) = \Sigma_{jj}(\omega)$, known as singular values, usually arranged in descending order. The expression for $x(\omega)$ becomes
\begin{eqnarray}
x(\omega) = V(\omega)\Sigma^{-1}(\omega)U(\omega)^{\dag}y(\omega),
\end{eqnarray}
$\mathbf{A}_{2}$ can be approximately decomposed into a summation of a few modes $v_{j}(\omega)$ corresponding to the smallest singular values as follows:
\begin{eqnarray}\label{Aexpansion}
\mathbf{A}_{2} \approx \sum_{j}\frac{u_{j}^{\dag}(\omega)y(\omega)}{\sigma_j^d(\omega)}\vec{\bf v}_{j}(\omega),
\end{eqnarray}
In this equation, the three components of $\vec{\bf v}_{j}$ along the three directions originate from the $3N$ elements of the vector $v_{j}(\omega)$. The term ${u_{j}^{\dag}(\omega)y(\omega)}/{\sigma_j^d(\omega)}$ represents the expansion coefficient of mode $\vec{\bf v}_{j}$.
In the context of the Lorentz gauge, the induced potential can also be approximated as:
\begin{eqnarray}
\phi_2(\omega) \approx \sum_{j}\frac{u_{j}^{\dag}(\omega)y(\omega)}{ik\epsilon_2\sigma_j^d(\omega)}(\nabla G_2) \cdot G_2^{-1}\vec{\bf v}_{j}(\omega).
\end{eqnarray}
The surface current $\bf h_2$ and surface charge $\sigma_2$ are decomposed as follow,
\begin{eqnarray}
{\bf h}_2(\omega) \approx \sum_{j}\frac{u_{j}^{\dag}(\omega)y(\omega)}{\sigma_j^d(\omega)}G_2^{-1}\vec{\bf v}_{j}(\omega),
\end{eqnarray}
and
\begin{eqnarray}
\sigma_2(\omega) \approx \sum_{j}\frac{u_{j}^{\dag}(\omega)y(\omega)}{ik\epsilon_2\sigma_j^d(\omega)}G_2^{-1}(\nabla G_2) \cdot G_2^{-1}\vec{\bf v}_{j}.
\end{eqnarray}
 We can similarly decompose the internal surface current and charge, denoted as $\mathbf{h}_1$ and $\sigma_1$, by exchanging the indices $1$ and $2$; the expansion coefficients may differ from those on the external side. However, to compute the scattering spectra, we only need the vector potential $\mathbf{A}_2$ and the mode $\mathbf{v}_j$.

It is possible that a global decomposition like Eq. (\ref{svd}) exists, where $M(\omega)$, $U(\omega)$, $\Sigma(\omega)$, and $V(\omega)$ are continuous matrix functions of the frequency $\omega$. The diagonal entries of $\Sigma(\omega)$ may not be arranged in descending order for all frequencies. Consequently, the mode $v_j(\omega)$ might shift as $\omega$ varies. In practice, the singular value decomposition in Eq. (\ref{svd}) has to be carried out for each specific frequency $\omega$. Then the establishment of such continuous matrix functions can be achieved through techniques like mode tracking \cite{safin_advanced_2016,suryadharma_singular-value_2017,suryadharma_quantifying_2019}.  

We employ the Galerkin approximation to circumvent the need for performing SVD at each frequency. Assuming that the variation with $\omega$ are gradual, we can leverage $U(\omega_0)$ and $V(\omega_0)$ at a chosen fixed frequency $\omega_0$ to approximate $U(\omega)$ and $V(\omega)$. Then $x(\omega)$ is expanded as a Galerkin approximation
\begin{eqnarray}\label{svd1}
     x(\omega) \approx  V(\omega_0)a(\omega) = \sum_j a_j(\omega) v_j(\omega_0),
  \end{eqnarray}
where $a_j(\omega)$ is the expansion coefficient and the element of the column vector $a(\omega)$. Subsequently, the vector $a(\omega)$ is solved using the following Galerkin equation :
\begin{eqnarray}\label{svd2}
\left[{U^{\dag}(\omega_0)}M(\omega)V(\omega_0)\right ]a(\omega)=U^{\dag}(\omega_0)y(\omega),\nonumber\\
a(\omega) = \left [ {U^{\dag}(\omega_0)}M(\omega)V(\omega_0)\right]^{-1} \left[U^{\dag}(\omega_0)y(\omega)\right],
\end{eqnarray}
which allows us to once again derive $a_j(\omega_0)={u_{j}^{\dag}(\omega_0)y(\omega_0)}/{\sigma_j^d(\omega_0)}$ when $\omega = \omega_0$. We still need to solve the equation at every frequency, but the dimension of $M$ can be reduced by transforming it into ${U^{\dag}(\omega_0)}M(\omega)V(\omega_0)$. In the case of nanoparticles with symmetric geometry, the rotation of a mode $v_j(\omega)$ with varying $\omega$ is anticipated to be confined within a subspace formed by a few vectors $v_j(\omega_0)$. This suggests that $v_j(\omega)$ can be approximated as a linear combination of a few $v_j(\omega_0)$.

\subsubsection{QNMs Decomposition}
In the second approach, we conduct the decomposition of the surface current through the utilization of natural modes or quasi-normal modes. These modes are positioned at the poles within the complex frequency plane. If we extend the frequency domain of the matrix functions as presented in Eqs. (\ref{CurrentMatrix}) and (\ref{svd}) from the real axis to encompass the complex plane, it is possible to identify a complex-valued frequency $\omega^{*}$ that satisfies
\begin{eqnarray}\label{natural_modes}
M(\omega^*)x(\omega^*)=0.
\end{eqnarray}
This is a nonlinear eigenvalue problem\cite{guttel_nonlinear_2017}. $x(\omega^*)$ corresponds to the natural resonance mode for $\mathbf{A}_{2}(\bf s)$ at resonance $\omega^*$. A distinctive characteristic of a natural mode is its ability to sustain itself even with zero external excitation ($y(\omega)$). Eq. (\ref{natural_modes}) implies the zero determinant of $M(\omega^*)$
\begin{eqnarray}
det\left[M(\omega^*)\right]=det\left[\Sigma(\omega^*)\right]= \prod_{i=1}^{3N} \sigma^d_i(\omega^{*}) = 0.
\end{eqnarray}
At a natural resonance frequency $\omega^{*}$, one or a few of the diagonal entries of 
$\Sigma(\omega^{*})$ become zero, i.e., $\sigma^d_j(\omega^{*}) = 0$. The number of zero diagonal diagonal entries of $\Sigma(\omega^*)$ usually corresponds to the degeneracy of that natural mode. The resonance frequency $\omega^*$ also corresponds to a pole of the inverse of the matrix function $M(\omega)$. Assuming $\frac{d\sigma^d_{j}(\omega)}{d\omega}\bigg|_{\omega = \omega^*} \neq 0$, we can obtain $M^{-1}$ near $\omega^*$ from Eq. (\ref{svd})
\begin{eqnarray}\label{keldysh}  
&M&^{-1}(\omega) = V(\omega)\Sigma^{-1}(\omega)U(\omega)^{\dag}= \nonumber\\
&\sum\limits_j&\left(\frac{d\sigma^d_{j}(\omega)}{d\omega}\bigg|_{\omega = \omega^*}(\omega - \omega^*)\right)^{-1}v_j(\omega^*)u_j(\omega^*)^{\dag} + R(\omega),\nonumber\\
\end{eqnarray}
where the meromorphic remainder $R(\omega)$ approaches zero at $\omega^*$, making it negligible for approximation purposes around $\omega^*$. The global form of Eq.~(\ref{keldysh}) contains a few poles $\omega^*_m$
\begin{align}\label{keldysh1}  
&M^{-1}(\omega) \nonumber\\
&= \sum\limits_{m,j}\left(\frac{d\sigma^d_{j}(\omega)}{d\omega}\bigg|_{\omega = \omega^*_m}(\omega - \omega^*_m)\right)^{-1}v_j(\omega^*_m)u_j(\omega^*_m)^{\dag} \nonumber\\&+ \tilde{R}(\omega).
\end{align}
The remainder $\tilde{R}(\omega)$ can be understood as a contour integration as 
\begin{eqnarray}\label{eq:remainder} 
&\tilde{R}(\omega) = \frac{1}{2\pi i}\oint_C \frac{M^{-1}(\omega_1) }{\omega_1-\omega}d\omega_1, 
\end{eqnarray}
where $C$ is the contour that encloses $\omega$ and the poles $\omega^*_m$ within its interior. If $M(\omega)$ is diagonalizable, Eq.~(\ref{keldysh}) can also be derived from the eigenvalue decomposition, where $\Sigma(\omega)$ is understood as the diagonal matrix of eigenvalues, and the columns of $U(\omega)$ and $V(\omega)$ are the left and right eigenvectors. Eq.~(\ref{keldysh1}) takes the form of Keldysh’s theorem for the case of semisimple eigenvalues \cite{guttel_nonlinear_2017}. The technique of contour integration can be used to identify the poles and the corresponding resonance modes, as detailed in Appendix~\ref{app:polefind}.

If the remainder is negligible, then $x(\omega)$ can be expressed as a series of $v_j(\omega^*_m)$:
\begin{align}\label{eq:simpleexpansion}
x(\omega) = \sum\limits_{m,j}\frac{\langle u_j(\omega^*_m) |y(\omega^*_m)\rangle}{\langle u_j(\omega^*_m) |M'(\omega^*_m) |v_j(\omega^*_m)\rangle}\frac{|v_j(\omega^*_m)\rangle}{\omega - \omega^*_m},
\end{align}
where $M'(\omega^*_m)$ represents the derivative of $M$ with respect to $\omega$ at $\omega^*_m$. The direct decomposition like Eq.~(\ref{eq:simpleexpansion}) is effective in some situations \cite{sauvan_theory_2013,powell_resonant_2014}. However, its accuracy can be inconsistent, which is influenced by factors such as the shape, size, and dielectric function of the nanoparticles \cite{unger_novel_2018}. Within the BEM framework, our examples illustrate that while Eq.~(\ref{eq:simpleexpansion}) captures the essential aspects of the optical response, there are discrepancies when compared to direct numerical calculations. This indicates a noteworthy contribution from the remainder term in our approach. Therefore, in lieu of direct decomposition, one might consider using the QNMs as an expansion basis set, as outlined in Eq.~(\ref{svd2}).

\section{\label{sec: dielectric} The dielectric function of gold}
\begin{figure}[hbt!]
\includegraphics[scale = 0.85]{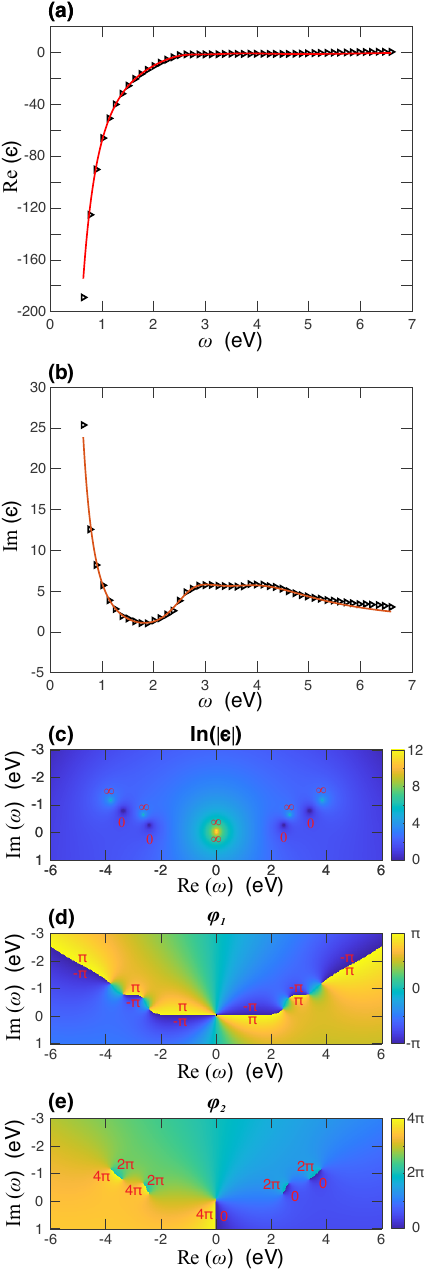}
\caption{\label{fig:gold_dielectric} The analytic dielectric function of gold. (a) and (b) Real and imaginary parts of the gold dielectric function. The data points (triangles) are from Johnson and Christy \cite{PhysRevB.6.4370}, while the solid red lines represent the analytical expression given by Eq.~(\ref{analyticM}). (c) Modulus of the gold dielectric function on the complex frequency plane, with the labels '0' and '$\infty$' indicating the values at zero and infinity, respectively. (d) and (e) Two choices of the phase and branch cuts, with the labels '0', '$\pm\pi$', '2$\pi$', and '4$\pi$' indicating the values on both sides of the branch cuts. The frequency unit is in eV. }
\end{figure}

The dielectric function of gold is often modeled using the Drude model, which is particularly effective in describing behavior at near-infrared frequencies. In this study, we adopt an analytical Drude-Lorentz model that extends the Drude approach by incorporating two interband transitions \cite{analyticmodel}. This enhanced model provides a more accurate representation of gold's dielectric function, especially within the ultraviolet (UV) and visible spectrum – a critical consideration for characterizing small gold nanoparticles. However, this choice introduces complexities in selecting an appropriate branch cut for refractive index determination, as detailed in the subsequent section. The Drude-Lorentz dielectric function relies on wavelength $\lambda$ (nm) given as follows,
\begin{align}\label{analyticM}
    \epsilon(\lambda) &= \epsilon_{\infty} - \frac{1}{\lambda_p^2(1/\lambda^2 + i/\gamma_p\lambda)}\nonumber\\
    &+ \sum_{j=1,2}\frac{A_j}{\lambda_j}\left[\frac{e^{i\phi_j}}{(1/\lambda_j-1/\lambda - i/\gamma_j)}\right.\nonumber\\
    &+ \left.\frac{e^{-i\phi_j}}{(1/\lambda_j+1/\lambda + i/\gamma_j)}\right],   
\end{align}
where the values of the parameters $A_j$, $\lambda_j$, $\gamma_j$, and $\phi_j$ are obtained through fitting the data of Johnson and Christy \cite{analyticmodel,PhysRevB.6.4370} to the model and listed in Table~\ref{table:gold_die}. The real and imaginary parts of the dielectric function are illustrated in Fig.~\ref{fig:gold_dielectric} (a) and (b), respectively. By utilizing this analytical model, we can obtain the dielectric function for complex frequency through the relation $\lambda = 2\pi c/\omega$, where $c$ is the speed of light. In Fig.~\ref{fig:gold_dielectric} (c), we illustrate the logarithm of $|\epsilon|$ on the complex frequency plane, highlighting the positions of zero and infinity values. 

\begin{table}[ht!]
\begin{ruledtabular}
\caption{Parameters of the analytic dielectric function of gold in Eq.~(\ref{analyticM}).}
\label{table:gold_die}
\centering
\begin{tabular}{c c c c c c} 
 
 $\epsilon_{\infty}$ & 1.54 &$A_1$& 1.27 & $A_2$& 1.1 \\  [0.5ex] 
 $\lambda_p$ (nm) &143 &$\phi_1$& $-\pi/4$ & $\phi_2$& $-\pi/4$ \\
 $\gamma_p$ (nm) &14500 &$\lambda_1$ (nm)& 470 & $\lambda_2$ (nm)& 325 \\
     & & $\gamma_1$ (nm)& 1900 & $\gamma_2$ (nm)& 1060 \\
\end{tabular}
\end{ruledtabular}
\end{table}

When representing the dielectric function as $\epsilon = |\epsilon|e^{i\phi}$, two mathematically permissible values arise for the refractive index $n = \sqrt{|\epsilon|}e^{i\frac{\phi}{2}}$. To ensure that the refractive index $n(\omega)$ remains a continuous single-valued function, two possible choices of the branch cut can be considered by defining the phase $\phi$ as $\phi_1$ and $\phi_2$. Fig.~\ref{fig:gold_dielectric} (d) and (e) illustrate the values of $\phi_1$ and $\phi_2$ on the complex plane of $\omega$, respectively. We use the second choice for the remaining part. More discussion can be found in Appendix~\ref{app:branch}. The refractive index jumps from negative to positive across the branch cut, leading to discontinuities in the matrix $M(\omega)$. The contour $C$ in Eq.~(\ref{eq:remainder}) should be chosen so that the branch cut lies outside the encircled region. There exist some exceptions, such as a sphere, where $M(\omega)$ remains continuous along the branch under spherical symmetry. However, this symmetry is compromised after surface discretization. Thus, in practice, we observe that $M(\omega)$ exhibits non-smooth behavior along certain portions of the branch cut for a gold nanosphere.

It should be noted that there is a relationship between the complex conjugate pair for both the dielectric function and the refractive index, given by $\overline{\epsilon(\omega)} = \epsilon(-\overline{\omega})$ and $\overline{n(\omega)} = n(-\overline{\omega})$. As a result, from Eq.(\ref{Mdefiniion}), one can deduce that $\overline{M(\omega)} = M(-\overline{\omega})$. If $\omega^*$ represents a natural frequency with an associated solution $x(\omega^*)$ such that $M(\omega^*)x(\omega^*) = 0$, then $-\overline{\omega^*}$ also stands as a natural frequency. The corresponding solution in this case is $\overline{x(\omega^*)}$, given that $M(-\overline{\omega^*})\overline{x(\omega^*)} = 0$. As substantiated in Appendix~\ref{app:leff_right}, $x(\omega^*)$ and $x(-\overline{\omega^*})$ constitute a pair of left and right eigenvectors, respectively.

\section{\label{sec:level3}Decomposition examples}
In this section, we apply the previously discussed method to analyze the scattering of incident electromagnetic plane waves by gold nanoparticles. Initially, we employ the decomposition techniques on both small and large spheres, which can be analytically studied using Mie theory. In these cases, the comparison to Mie theory allows us to benchmark these decomposition methods. Subsequently, we apply these techniques to more complex scenarios, considering cases such as sphere clusters and triangles.

\subsection{Small Gold Sphere}
\begin{figure}
\includegraphics[scale = 1.2]{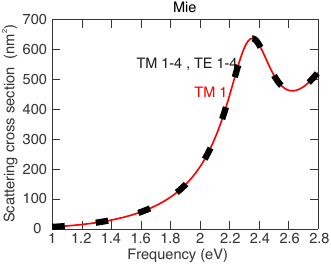}
\caption{\label{fig1_mie} Scattering cross section calculated from Mie theory for a 64 nm diameter gold sphere embedded in air (refractive index $n=1$). The modes up to 4th-order TM and TE modes (black dashed line) and only the 1st-order TM mode (red solid line) exhibit identical spectra.}
\end{figure}

We commence by examining the scattering of a plane x-polarized wave by a gold sphere with a diameter of 64 nm in air (refractive index 1). Mie theory offers an analytical solution for this scattering problem and can serve as a benchmark to evaluate the goodness of the mode decomposition. For a sphere in the uniform background, vector spherical harmonics are the normal mode solutions of the Helmholtz equation without external source in spherical coordinates. Both the incident and scattering fields can be expressed through an expansion involving vector spherical harmonics, denoted by TM modes and TE modes. See details of Mie theory in Appendix~\ref{app:mie} and Ref.~\cite{bohren_absorption_1998}. Fig. \ref{fig1_mie} show total scattering cross section of the sum of  TE and TM modes from the first order to the fourth order and the scattering contribution of the dipolar mode, TM 1 (the first order). The scattering is mainly contributed by the first-order TM mode, TM 1, and other higher order modes are negligible. 

\begin{figure}
\includegraphics[scale = 1.2]{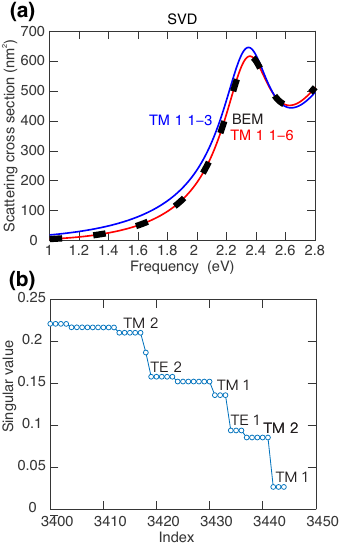}
\caption{\label{fig1_svd} Scattering cross section calculated from BEM and constructed from the SVD modes for a 64 nm diameter gold sphere embedded in air (refractive index $n=1$). (a) The BEM simulation (black dashed line) and the six 1st-order TM modes (TM 1 1-6 (SVD), red solid line) obtained from SVD yield the identical scattering cross section spectra. The scattering cross section built from TM 1 1-3 (SVD) is shown as a blue solid line. (b) Singular values.}
\end{figure}

\begin{figure*}
\includegraphics[scale = 1.2]{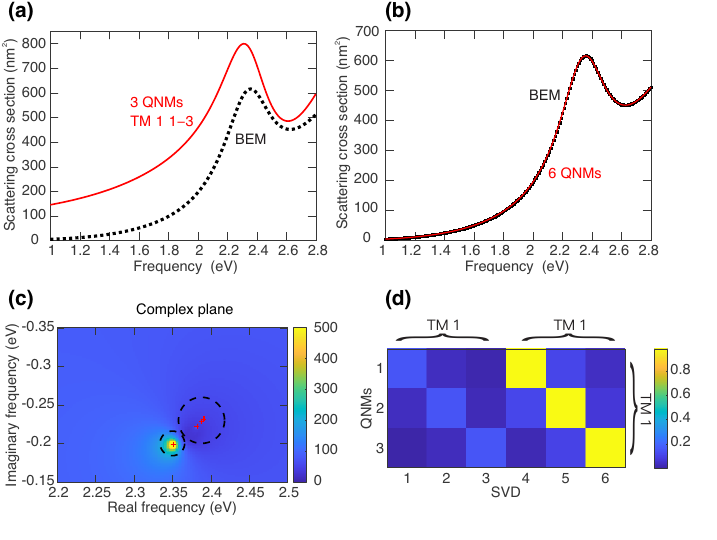}
\caption{\label{fig1_qnm_gk} Scattering cross section calculated from BEM and constructed from the QNMs for a 64 nm diameter gold sphere embedded in air (refractive index $n=1$). (a) Scattering cross section spectra calculated by BEM (black dotted line) and constructed by the three QNMs (TM 1 1-3 (QNMs)) (red solid line). (b) Scattering cross section spectra calculated by BEM (black dotted line) and constructed by the nine QNMs (three sets of TM 1 1-3 (QNMs)) (red solid line). (c) Complex map of condition number of $M(\omega)$. The two dashed lines represent the two contours to find the resonances and the natural modes within the circles. The red crosses indicates the locations of the three modes of TM 1 (QNMs).  (d) Inner products of three QNMs and six SVD modes, $v_{m}(\omega^*_m)$ and $v_{m}(\omega_0)$.}
\end{figure*}

Fig. \ref{fig1_svd} (a) shows the scattering cross sections calculated by BEM and those constructed by using the SVD modes as a basis set. In the BEM simulation, the surface of the sphere is discretized into meshes consisting of 1148 triangles. Therefore, $M(\omega)$ in Eq. (\ref{Mdefiniion}) is a $3444\times3444$ matrix; At $\omega = 2.0$ eV, $M(\omega)$ is decomposed through SVD, and its smallest singular values are plotted in Fig. \ref{fig1_svd} (b). The left and right singular vectors are used as a basis set for the Galerkin approximation according to Eq.~(\ref{svd2}).  By comparing to the cross section calculated by Mie theory, we categorize the modes from SVDs into TM and TE modes. There are six modes of TM 1 (SVD), three of which have the smallest singular values. These six modes yield nearly identical scattering cross section, as calculated directly by BEM. However, if only using the three TM 1 modes with the lowest singular values, the constructed scattering cross section (blue solid line in Fig. \ref{fig1_svd} (a) ) deviates significantly from the outcomes of BEM and the six TM 1 modes. TM 2 (SVD) modes also have relatively small singular values; nonetheless, their contribution to the scattering remains negligible. This is because the plane wave cannot efficiently excite TM 2 (SVD) as due to the small value of $u_{j}^{\dag}(\omega)y(\omega)$ in Eq. (\ref{Aexpansion}); also, the quadrupolar radiation from the TM 2 mode is much less efficient than the dipolar radiation from the TM 1 modes.

Fig. \ref{fig1_qnm_gk}(a) displays the scattering cross section computed using BEM and reconstructed employing QNMs as a basis set. The frequency locations of the QNMs are depicted in the complex frequency plane in Fig. \ref{fig1_qnm_gk}(c). The false-color map illustrates the condition number of $M(\omega)$ over the complex plane of the frequency $\omega$. The hotspots in the map correspond to the poles of $M^{-1}(\omega)$, where the resonances (red crosses) of the natural modes or QNMs can be found by using the contour integration. The three degenerate modes at the complex resonance of $2.35 - 0.20i$ eV, referred as TM 1 (QNMs), are employed as a set of Galerkin  basis according to Eq.~(\ref{pole2}), and then the corresponding scattering cross section can be calculated as shown in Fig. \ref{fig1_qnm_gk} (a). Although these three TM 1 modes (QNMs) capture the primary scattering characteristics, they do not reproduce the outcomes from BEM, unlike the construction from SVD modes. The root of this disparity can be inferred from the absolute inner products between the two sets of TM modes, six SVD modes and three QNMs, which are depicted in Fig. \ref{fig1_qnm_gk}(d). As underscored by the pronounced values of the absolute inner products, the three TM 1 modes from the QNMs bear a strong resemblance to the three TM 1 modes from the SVD possessing the smallest singular values. However, the other three pivotal SVD modes are not well represented. To reproduce the scattering, we need to find the other three proper QNMs into the basis set. As discussed in Appendix~\ref{app:mie}, the vector potential of the TM modes at any frequency can be expressed in a combination of the vector potentials of the TM modes at any other two frequency. Notably, the TM 1 modes (QNMs) can also be found at other frequencies. Nine TM1 (QNMs), located at $2.35-0.20i$ eV, $3.21-0.76i$ eV, and $4.59 - 1.67i$ eV, are used as the Galerkin basis set.  This allows for a more precise replication of the scattering cross section as shown in Fig. \ref{fig1_qnm_gk} (d). For analogous reasons, it is worth noting that a distinct expansion approach for the electric and magnetic fields has been proposed in Ref.~\cite{unger_novel_2018}.  In theory, the scattering spectra for nanoparticles of any shape can be precisely constructed if an adequate number of appropriate QNMs are employed. However, these QNMs typically lie near the pole and branch cut in the complex frequency plane, making the algorithms used to identify them potentially unstable.

\begin{figure}
\includegraphics{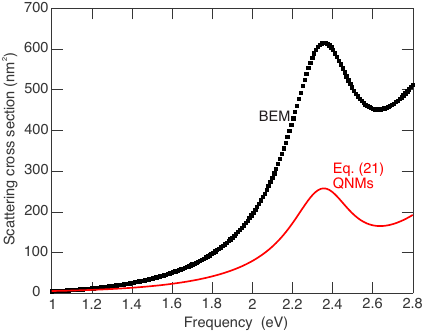}
\caption{\label{qnm_simple}Scattering cross section for a 64 nm diameter gold sphere: direct BEM simulation (black squares) vs. simple QNMs expansion from Eq.(\ref{eq:simpleexpansion}) (red solid line).}
\end{figure}

We further assess the effectiveness of the simple decomposition using Eq.~(\ref{eq:simpleexpansion}). Figure \ref{qnm_simple} illustrates the utilization of six sets of triply degenerate modes at energies of $\pm2.35-0.20i$ eV, $\pm3.21-0.76i$ eV, and $\pm4.59 - 1.67i$ eV. A noticeable difference exists between the scattering constructed from the simple decomposition and the result obtained directly through the BEM simulation. This discrepancy indicates that the remainder term, Eq.~(\ref{eq:remainder}, is not negligible. The convergence of QNMs construction with respect to the direct simulation varies among different studies \cite{sauvan_theory_2013,lalanne_light_2018,powell_interference_2017,unger_novel_2018,lasson_three-dimensional_2013}. Such variation depends on factors like the dielectric function, the size and geometry of the simulated nanostructures, and the formalism of the equations employed.  Different formulations, like unique integral equations, lead to disparate matrices $M$ post-discretization. This means that the choice of matrix $M$ is not unique. Conversely, it is feasible to redefine a new matrix $M^{new}$ and $Y^{new}$ to replace $M$ and $Y$, such that the new remainder term $\frac{1}{2\pi i}\oint_C \frac{(M^{new})^{-1}(\omega_1) }{\omega_1-\omega}d\omega_1$ is minimized. Then it enables a better approximation of the new matrix $M^{new}$ through a new set QNMs using Eq.~(\ref{eq:simpleexpansion}). We are currently exploring this possibility, although it lies beyond the scope of the present study. For the subsequent examples, we will exclusively employ QNMs as a basis set for the Galerkin approximation.

\begin{figure*}
\includegraphics{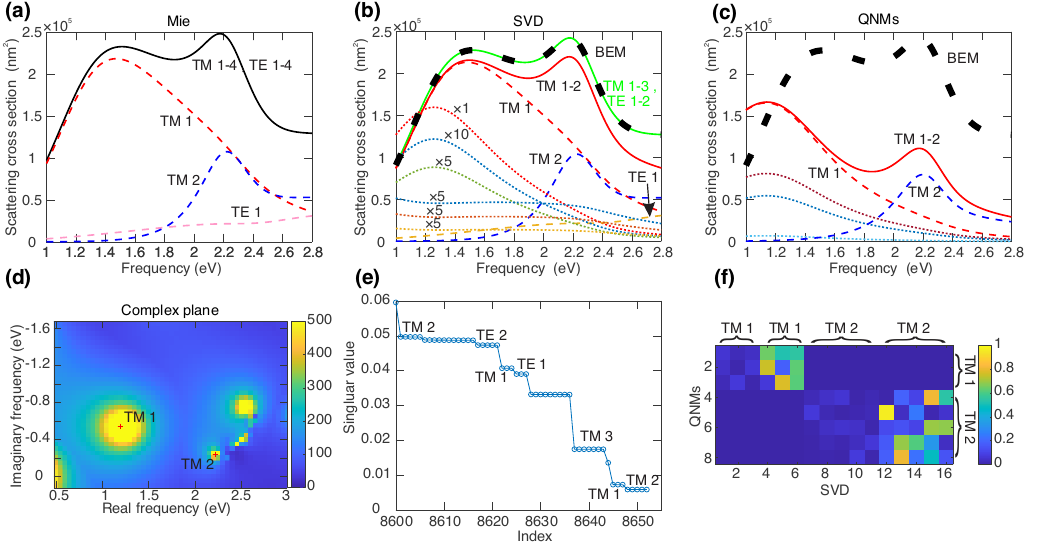}
\caption{\label{bigsphere}Scattering of a 300 nm diameter gold sphere embedded in air (refractive index $n=1$). (a) Scattering cross section calculated from Mie theory. The black solid line represents the scattering cross section based on modes up to the 4th-order TM and TE modes (TM 1-4 and TE 1-4). Also shown are TM 1 (red dashed line), TM 2 (blue dashed line), and TE 1 (yellow dashed line). (b) The BEM simulation (black dashed line) and the modes obtained from SVD including TM 1-3 and TE 1-2 (green solid line) yield the identical scattering cross section spectra. The scattering cross section of TM 1 (SVD), and TM 2 (SVD), TE 1 (SVD) are represented by red, blue, and yellow dashed lines, respectively. The red solid line depicts the total scattering of TM 1 (SVD) and TM 2 (SVD). Dot lines indicate the six TM 1 (SVD) modes (magnified by 1-10 times for clarity). (c) Scattering cross section spectra calculated by BEM (black dashed line) and constructed from the eight QNMs (TM 1-2) (red solid line). The scattering cross section of TM 1 (QNMs) and TM 2 (QNMs) are shown as red and blue dashed lines, respectively. Dot lines represent the three TM 1 (QNMs) modes. (d) Complex map of condition number of $M(\omega)$. The red crosses indicate the locations of TM 1 (QNMs) and TM 2 (QNMs). (e) Singular values.  (f) Inner products of eight QNMs and sixteen SVD modes, $v_{m}(\omega^*_m)$ and $v_{m}(\omega_0)$.}
\end{figure*}

\begin{figure*}
\includegraphics{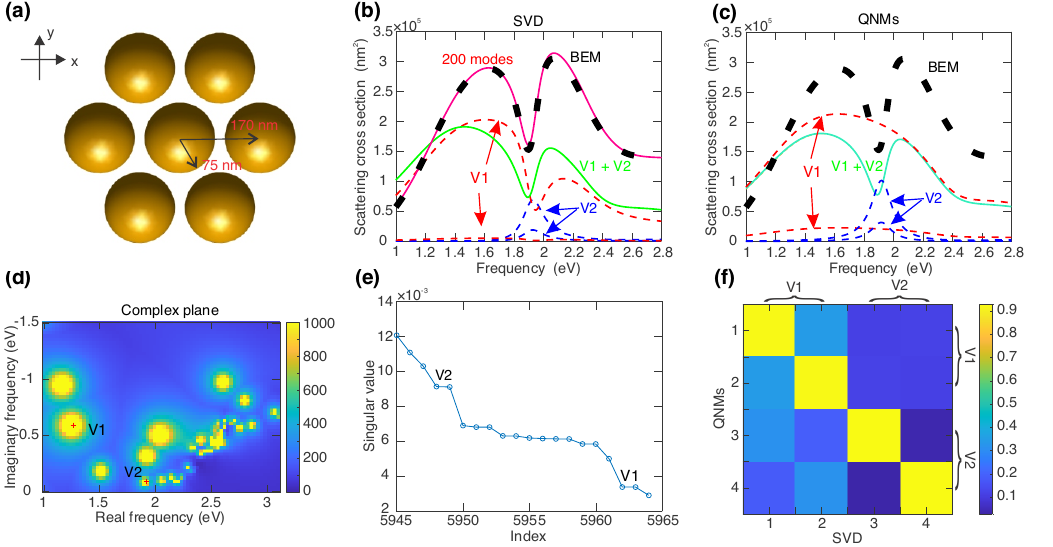}
\caption{\label{fig3} Scattering of a heptamer system. (a) The heptamer structure comprises seven gold spheres with a radius of 75 nm, arranged at a neighboring distance of 170 nm. (b) Scattering cross section spectra calculated by BEM (black dashed line), constructed by the 200 modes from SVD (red solid line) and constructed by the combined V1 (SVD) and V2 (SVD) modes (green solid line).  The red and blue dash lines are the scattering cross section built from V1 (SVD) and V2 (SVD), respectively. (c) Scattering cross section spectra calculated by BEM (black dashed line) and constructed by the combined V1 (QNMs) and V2 (QNMs) modes (green solid line).  The red and blue dash lines are the scattering cross section built from V1 (QNMs) and V2 (QNMs), respectively. (d) Complex map of condition number of $M(\omega)$. The red crosses indicate the locations of V1 (QNMs) and V2 (QNMs).  (e) Singular values.  (f) Inner products of four QNMs and four SVD modes, $v_{m}(\omega^*_m)$ and $v_{m}(\omega_0)$.}
\end{figure*}

\begin{figure*}
\includegraphics{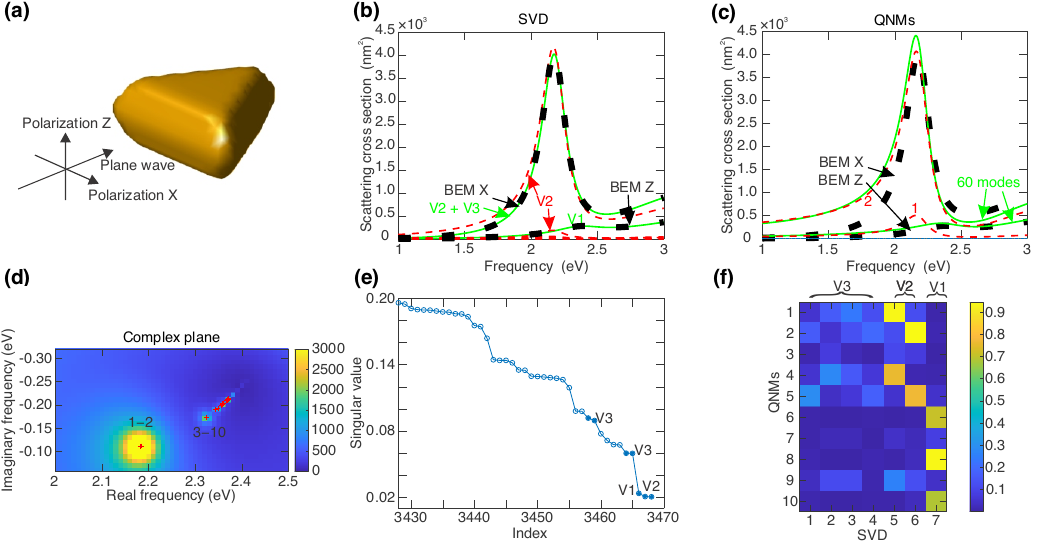}
\caption{\label{fig4} Scattering of a gold nanotriangle. (a) Gold nanotriangle. X and Z polarized plane waves are used for excitaton. (b) Scattering cross sections for X and Y polarizations are calculated by BEM (black dashed lines) and  constructed with V1 (SVD), V2 (SVD) and V3 (SVD) modes (green solid line). The red dash lines indicate the scattering cross section independently constructed from V2 and V3 modes. (c) Scattering cross sections for X and Y polarizations are calculated by BEM (black dashed lines) and constructed with 60 QNMs (green solid lines). The dash lines represent the scattering cross section separately constructed from QNM 1-2. (d) Complex map of condition number of $M(\omega)$. The red crosses indicate the locations of QNMs.  (e) Singular values.  (f) Inner products of ten QNMs and seven SVD modes, $v_{m}(\omega^*_m)$ and $v_{m}(\omega_0)$.}
\end{figure*}

\subsection{Big Gold Sphere}
While the scattering is predominantly driven by the TM 1 modes for the small gold sphere, the influence of higher-order modes becomes increasingly significant as the gold sphere size increases. Figure \ref{bigsphere}(a) demonstrates noteworthy contributions from TM 1, TM 2, and TE 1 modes for a gold sphere with a diameter of 300 nm, as evidenced by the analysis of Mie scattering. In the BEM calculation, we discretize the surface of this larger gold sphere into 2884 triangles. After performing singular value decomposition on the $8652\times8652$ matrix $M(\omega)$ at the frequency $\omega = 1.91$ eV, the smallest singular values are plotted in Figure \ref{bigsphere}(e), some being identified as TE and TM modes. The scattering cross sections are plotted for six TM 1 (SVD) modes, three TE 1 (SVD) modes, and ten TM 1 (SVD) modes in Figure \ref{bigsphere}(b). These modes accurately replicate the scattering cross sections of the corresponding modes calculated by Mie theory. Notably, although modes based on the vector potential are orthogonal on the surface by definition in SVD, the calculated electric and magnetic fields from these orthogonal vector potential modes might not be orthogonal in the entire space. Consequently, the total TM 1 (SVD) scattering intensity cannot be simply computed as the sum of the scattering intensity of each TM 1 (SVD) mode. However, the electric and magnetic fields calculated from TM 1 (SVD) are still orthogonal to those of TM 2 (SVD). Therefore, once the scattering intensities are obtained for both TM 1 (SVD) and TM 2 (SVD) modes, their combined scattering intensity can be computed as the sum of their respective intensities.

The mapping of the condition number of $M(\omega)$ with respect to complex-valued $\omega$ is illustrated in Fig. \ref{bigsphere}(d). By utilizing contour integration, we have successfully determined three TM 1 (QNMs) modes and five TM 2 (QNMs) modes. The corresponding scattering spectra for these modes are shown in Fig. \ref{bigsphere}(c). A noticeable disparity emerges between the TM 1-2 (QNMs) and BEM simulation results. This discrepancy arises due to the fact that the set of eight TM 1-2 (QNMs) modes only constitutes half of the mode set of TM 1-2 (SVD). This distinction is highlighted by the inner product displayed in Fig. \ref{bigsphere}(f) between these two mode sets. We can introduce other TM 1-2 (QNMs) into the basis set to give a more accurate result.

\subsection{Assembly of Gold Spheres}
 Mie theory elegantly resolves the scattering problem for the aforementioned sphere examples. However, applying it directly to intricate assembly structures involving multiple gold nanoparticles might not be as straightforward as in the previous examples. The plasmonic phenomena, including dark states and Fano resonances,  are not only reliant on individual nanoparticles but also hinges on the intricate interaction between them \cite{nordlander_plasmon_2004,halas_plasmons_2011}. This complexity is exemplified by the gold heptamer showcased in Fig. \ref{fig3}(a), consisting of seven gold spheres. In the BEM simulation, each gold sphere's radius is set as 75 nm, and the distance is set as 170 nm from the center of one sphere to the center of its nearest neighbor. The background refractive index is set at 1. A similar structure can be found in the early work \cite{hentschel_transition_2010}. For the BEM calculation, we mesh each sphere's surface into 284 triangles.  For a X-polarized incident plane wave,  the calculated scattering cross section is shown as the black dashed line in Fig. \ref{fig3} (b) and (c). A distinct Fano resonance manifests as a dip around 1.9 eV, stemming from the destructive interference between the subradiant and supperradiant modes.  SVD is performed on the $5964 \times 5964$ matrix function $M(\omega)$ at the complex frequency $\omega$ of $1.26 - 0.59i$ eV. The corresponding smallest singular values are detailed in Fig.~\ref{fig3}(e). In Fig.~\ref{fig3}(b), the scattering cross section can be effectively reproduced using 200 modes obtained from SVD in the Galerkin approximation. Among these modes, four modes designated as V1 and V2 play a pivotal role in shaping the primary scattering characteristics. Notably, both of these modes are doubly degenerate due to their two polarizations. V1 (SVD) has a dip around 1.95 eV in its scattering cross section. It further constructively and destructively interacts with V2 (SVD), shifting the dip to 1.90 eV as shown in the combined scattering cross section of V1 (SVD) and V2 (SVD).  
 
 In Fig.~\ref{fig3}(d), we identify four QNMs, denoted as V1 (QNMs) and V2 (QNMs), positioned at the poles of the complex plane of $M(\omega)$. Although incorporating more QNMs can enhance the accuracy of the constructed spectra, identifying the appropriate modes near the branch cut using the contour-based algorithm is computationally intensive. Remarkably, these four QNMs correspond to V1 and V2 from the SVD, validated by the inner products between these mode sets in Fig. \ref{fig3}(f).   Utilizing these four QNMs as a basis, we compute a comparable scattering cross section. However, it is important to note that their individual scattering cross sections differ from those of V1 (SVD) and V2 (SVD), particularly in the absence of a dip within the scattering cross sections of V1 (QNMs). The dip observed in the collective scattering of V1 (QNMs) and V2 (QNMs) arises solely from their destructive interaction. Fig.~\ref{fig3}(f) indicates that V1 (SVD) is a blend of V1 (QNMs) and V2 (QNMs), resulting in a dip in the scattering cross section of V1 (SVD).  It is worth noting that the contrast between modes defined by eigenvalue and singular value problems has been explored in prior research \cite{suryadharma_singular-value_2017,suryadharma_quantifying_2019}.

\subsection{Gold Nanotriangle}
In our final example, we apply the decomposition methods to a gold nanotriangle. Such structures can be made either through wet synthesis or by nanofabrication. They manifest a collection of dense modes, which have been investigated extensively, especially through electron beam excitation \cite{nelayah_mapping_2007,das_probing_2012,myroshnychenko_unveiling_2018}.  As shown in Fig. \ref{fig4}(a), our simulation involves a gold nanotriangle in the shape of a truncated triangular bipyramid. In simluation, the geometric model is built by truncating two tetrahedrons with 104 nm edges. Post-truncation, the nanotriangle has a thickness of 40 nm. Fig. \ref{fig4}(a) indicates the use of X-polarized and Z-polarized incident plane waves as the excitation source. To perform the BEM calculation, the nanotriangle's surface is discretized into 1156 triangles. The calculated scattering cross sections for two polarizations are represented by the black dashed solid curves in panels (b) and (c) of Fig.~\ref{fig4}.  The smallest singular values of the $3468 \time 3468$ matrix $M(\omega)$ with $\omega$ at 2.20 eV are presented in Fig. \ref{fig4}(e). While sixty modes with the lowest singular values are used in Eq. (\ref{Aexpansion}), our focus narrows down to seven specific modes labeled V1-V3 (SVD) as demonstrated by the scattering spectra in 
Fig.~\ref{fig4}(b). In the case of the X polarization, the scattering cross section calculated by the BEM can be accurately reproduced by V2 (SVD) and V3 (SVD); for the Z polarization, the scattering is primarily governed by the mode V1 (SVD). 

We also check the expansion using QNMs as a basis set, as depicted in Fig. \ref{fig4} (c) and (d). Sixty QNMs, whose resonances are in the range of the map in panel (d),  are employed to reconstruct the scattering cross sections, represented by the green solid lines in Fig. \ref{fig4}(c). However, there is a notable deviation from the BEM and SVD results, because these sixty QNMs are insufficient to project V1-V3 (SVD).

\section{\label{sec:level4}Conclusion}
In this study, we thoroughly investigated mode decomposition strategies within the boundary element method (BEM), focusing on singular value decomposition (SVD) and quasi-normal mode (QNM) decomposition. These approaches enabled us to capture the characteristic plasmonic responses from a few fundamental modes. While QNMs provide a deeper understanding of parameters such as frequency and damping, they are challenging to identify and utilize for spectral reconstruction, especially in large and complex structures. By integrating SVD modes into a Galerkin basis set, we achieved a balance between computational efficiency and modal expansion accuracy. This combination of BEM with mode decomposition methods enhances its suitability for the rapid understanding and design of individual plasmonic modes for specific applications.

\appendix
\section{A brief review of BEM}\label{app:BEM}
 The potential-based boundary element method (BEM) \cite{garcia_de_abajo_retarded_2002,RevModPhys.82.209,nelayah_mapping_2007,polman_electron-beam_2019} is a widely used approach for solving Maxwell's equations in the presence of arbitrarily shaped metallic nanoparticles. It has been implemented in the open-source MATLAB toolbox MNPBEM \cite{hohenester_mnpbem_2012,waxenegger_plasmonics_2015}. In the BEM, the total electric and magnetic fields are represented in terms of the total scalar and vector potentials $\phi^{tot}({\bf r})$ and ${\bf A}^{tot}({\bf r})$:
\begin{equation}
{\bf E^{tot}} = ik {\bf A}^{tot} - \nabla \phi^{tot}, \quad \textrm{and} \quad {\bf H}^{tot} = \frac{1}{\mu}\nabla \times {\bf A}^{tot},
\end{equation}
with the Lorentz gauge
\begin{equation}
    \nabla \cdot {\bf A^{tot}} = ik\epsilon\mu \phi^{tot}.
\end{equation}
Here the wavenumber $k = \omega /c$, and $c$ is the speed of light. As shown by Fig.~\ref{fig:BEM}, the abrupt boundary of a nanoparticle divides space into an inside region $V_1$ and an outside region $V_2$. The dielectric function $\epsilon({\bf r},\omega)$ and magnetic permeability $\mu({\bf r}, \omega)$ are assumed to be frequency-dependent, local, and isotropic. While these properties remain uniform within each region, they may undergo a sharp change across the boundary. 

The general forms of the total scalar and vector potentials in each region are expressed into the external and induced parts
\begin{align}
    \phi^{tot}({\bf r}) &= \int_{S}d{\bf s}G_{1,2}(|{\bf r - s|})\sigma_{1,2}({\bf s}) + \phi_{1,2}^e({\bf r}) \\
    &= \phi_{1,2} + \phi_{1,2}^e
\end{align}
and 
\begin{align}
    {\bf A}^{tot}(\bf r) &= \int_{S}d{\bf s}G_{1,2}(|{\bf r - s}|){\bf h}_{1,2}({\bf s}) + {\bf A}_{1,2}^e({\bf r}) \\
    &= {\bf A}_{1,2} + {\bf A}_{1,2}^e.
\end{align}
Here $\phi_{1,2}^e$ and $\mathbf{A}_{1,2}^e$ represent the external parts, namely, the scalar and vector potentials that are generated by external perturbation or excitation from the region $j=1,2$, assuming the entire space is homogeneous with $\epsilon_{1,2}$ and $\mu_{1,2}$. The induced parts of scalar and vector potentials, $\phi_{1,2}$ and $\mathbf{A}_{1,2}$, can be mimicked through boundary integrals of surface charge $\sigma_{1,2}$ and the surface current $\bf h_{1,2}$ along the surface infinitesimally close to the boundary from the two sides. The integrals are denoted as $G_{1,2}\sigma_{1,2}$ and $G_{1,2}\mathbf{h}_{1,2}$. $G_{1,2}$ is the Green function of the wave equation
\begin{equation}
    [\nabla^2 + k_{1,2}^2]G_{1,2}( r)=-4\pi \delta ({\bf r}) \quad \textrm{and} \quad G_{1,2}( r)=\frac{e^{ik_{1,2}r}}{r},
\end{equation}
where $k_{1,2} = k\sqrt{\epsilon_{1,2} \mu_{1,2}}$. 

The surface charges $\sigma_1(\bf s)$ and $\sigma_2(\bf s)$, and the surface currents ${\bf h}_1(\bf s)$ and ${\bf h}_2(\bf s)$,  on each side of the boundary, are chosen so that the boundary conditions of electric and magnetic fields are satisfied. With the above equations, for nonmagnetic case such as gold here ($\mu_{1,2} = 1$), the outgoing solution of the Maxwell's equations, which vanishes at infinity, can be alternatively obtained from the four equations based on the scalar and vector potentials $\phi(\bf r)_{1,2}$  and ${\bf A(r)}_{1,2}$ \cite{garcia_de_abajo_retarded_2002},
\begin{equation}\label{foureqs_1}
    G_1\sigma_1 - G_2\sigma_2 = \phi_2^e - \phi_1^e,
\end{equation}
\begin{equation}\label{foureqs_2}
    G_1{\bf h}_1 - G_2{\bf h}_2 = {\bf A}_2^e - {\bf A}_1^e,
\end{equation}
\begin{equation}\label{foureqs_3}
H_1{\bf h}_1 - H_2{\bf h}_2 - ik{\bf n}_s(G_1\epsilon_1\sigma_1 - G_2\epsilon_2\sigma_2) = \vec{\alpha},
\end{equation}
\begin{equation}\label{foureqs_4}
H_1\epsilon_1\sigma_1 - H_2\epsilon_2\sigma_2 - ik{\bf n}_s\cdot(G_1\epsilon_1{\bf h}_1 - G_2\epsilon_2{\bf h}_2) = D^e,
\end{equation}
where
\begin{equation*}
\vec{\alpha} = {\bf n}_s\cdot\nabla({\bf A}_2^e - {\bf A}_1^e) + ik{\bf n}_s(\epsilon_1\phi_1^e - \epsilon_2\phi_2^e)
\end{equation*}
and 
\begin{equation*}
D^e = {\bf n}_s\cdot[\epsilon_1(ik{\bf A}_1^e - \nabla\phi_1^e)- \epsilon_2(ik{\bf A}_2^e - \nabla\phi_2^e)].
\end{equation*}
${\bf n}_s$ is the surface normal of the boundary at $s$, directed from 1 to 2; $H_{1,2}$ is the surface derivative of the green function $G_{1,2}$ on each side of the interface as $H_{1,2} = {\bf n}_s\cdot\nabla G_{1,2}(s,s') \pm 2\pi \delta(s,s')$.

Eqs. (\ref{foureqs_1}) - (\ref{foureqs_4}) can be solved numerically by discretizing the boundary into $N$ pieces. In this way, the operators $G_{1,2}$ and $H_{1,2}$ becomes $N\times N$ matrices, and Eqs.~(\ref{foureqs_1}) - (\ref{foureqs_4}) are transformed into a set of $8N$ linear equations. The solution $(\sigma_{1},\sigma_{2},{\bf h}_1,{\bf h}_2)$ becomes an $8N$-dimensional complex vector. 

To reduce the computation cost, one can derive Eqs.~(\ref{foureqs}) after some algebra manipulations, and then solve $(\sigma_{2},{\bf h}_2)$ from $4N$ linear equations. Alternatively, one can also initially solve for $\sigma_{2}$ from the following equation: 
\begin{eqnarray}
[\Gamma + k^2(L_1-L_2){\bf n }_s \cdot \Delta ^{-1} {\bf n }_s(L_1-L_2)]G_2\sigma_2\nonumber\\
= {D}^{e*} + ik(L_1-L_2){\bf n }_s\cdot\Delta ^{-1}\vec{ \alpha}^{*},
\end{eqnarray}
where
\begin{equation}
L_{j} = G_{j}\epsilon_{j}G_{j}^{-1},  \quad \Sigma_j = H_jG_j^{-1} \quad \textrm{with $j = 1,2$,}
\end{equation}
\begin{equation}
\Delta = \Sigma_1 - \Sigma_2,   \quad \Gamma = \epsilon_1\Sigma_1 - \epsilon_2\Sigma_2,
\end{equation}
and
\begin{eqnarray}
&{D}^{e*} = {D}^{e} - \epsilon_1\Sigma_1(\phi^e_2-\phi^e_1)+ik{\bf n}_s\cdot L_1({\bf A}_2^e-{\bf A}_1^e),\nonumber\\
&\vec{ \alpha}^* = \vec{ \alpha} - \Sigma_1({\bf A}_2^e-{\bf A}_1^e)+ik{\bf n}_sL_1(\phi^e_2-\phi^e_1).
\end{eqnarray}
Once $\sigma_2$ is solved, we get ${\bf h}_2$ from
\begin{equation}
\Delta G_2 {\bf h}_2= \vec{ \alpha}^* + ik(L_1-L_2){\bf n}_sG_2\sigma_2.
\end{equation}
By substituting $\sigma_2$ and ${\bf h}_2$ into  Eqs. (\ref{foureqs_1}) and (\ref{foureqs_2}), we obtain $\sigma_1$ and ${\bf h}_1$. Then electric and magnetic fields can be computed from  $\sigma_{1,2}$ and ${\bf h}_{1,2}$. The details of the derivation above can be found in the work of ~\citet{garcia_de_abajo_retarded_2002}.

In this paper, we instead solve for the $3N$-dimensional ${\bf h}_2$ first, using a set of $3N$ linear equations from 
\begin{eqnarray}\label{current_equations1}
[\Delta + k^2(L_1-L_2){\bf n }_s \Gamma ^{-1} {\bf n }_s\cdot(L_1-L_2)]G_2{\bf h }_2\nonumber\\
= \vec{ \alpha}^{*} + ik(L_1-L_2){\bf n }_s\Gamma ^{-1}{D}^{e*}. 
\end{eqnarray}
Subsequently, we determine $\sigma_2$ from 
\begin{eqnarray}
\Gamma G_2\sigma_2 = D^{e*} + ik(L_1-L_2){\bf n}_s\cdot G_2 {\bf h}_2.
\end{eqnarray}

\section{Solver based on contour integral}\label{app:polefind}
The method of contour integration can be employed to locate the poles and their corresponding resonance modes. Suppose that a collection of natural resonance frequencies $\omega^*$ exist in a region enclosed by a closed contour $C$; for each resonance $\omega^*_m$, including the case of degeneracy, there exists a eigen triplet of $(v_{m}(\omega^*_m),u_{m}(\omega^*_m),\sigma^d_{m}=0)$.  Cauchy's integral theorem is applied to two contour integrals:
\begin{widetext}
\begin{eqnarray}
A_0 &\overset{\mathrm{def}}{=} &\frac{1}{2\pi i}\oint_C M^{-1}(\omega) d\omega = \sum_{m}\left(\frac{d\sigma^d_{m}(\omega)}{d\omega}\bigg|_{\omega = \omega^*_{m}}\right)^{-1}v_{m}(\omega^*_{m})u_{m}^{\dag}(\omega^*_{m})=\tilde{V}\tilde{J}\tilde{U}^{\dag} \overset{\mathrm{SVD}}{=} \tilde{V}_0\tilde{J}_0\tilde{U}_0^{\dag},\label{A_0}\\
A_1 &\overset{\mathrm{def}}{=} &\frac{1}{2\pi i}\oint_C M^{-1}(\omega)\omega d\omega = \sum_{m}\left(\frac{d\sigma^d_{m}(\omega)}{d\omega}\bigg|_{\omega = \omega^*_{m}}\right)^{-1}\omega^*_{m}v_{m}(\omega^*_{m})u_{m}^{\dag}(\omega^*_{m})=\tilde{V}\tilde{O}\tilde{J}\tilde{U}^{\dag},
\end{eqnarray}
\end{widetext}
where the columns of $\tilde{V}$ and $\tilde{U}$ are $v_{m}(\omega^*_m)$ and $u_{m}(\omega^*_m)$, respectively; $\tilde{J}$ and $\tilde{O}$ are diagonal matrices, of which the diagonal entries are $\left(\frac{d\sigma^d_{m}(\omega)}{d\omega}\bigg|_{\omega = \omega^*_{m}}\right)^{-1}$ and $\omega^*_{m}$, respectively. The right end of Eq. (\ref{A_0}) represents the singluar value decomposition (SVD) of $A_0$. Consequently, we find 
\begin{equation}
\tilde{J}\tilde{U}^{\dag} = \tilde{V}^{-1}\tilde{V}_0\tilde{J}_0\tilde{U}_0^{\dag},
\end{equation}
and from this we define
\begin{eqnarray}
A_2 \overset{\mathrm{def}}{=} \tilde{V}_0^{\dag}A_1\tilde{U}_0 \tilde{J}_0^{-1}= \tilde{V}_0^{\dag}\tilde{V}\tilde{O}\tilde{J}\tilde{U}^{\dag} \tilde{U}_0 \tilde{J}_0^{-1}\\
=\tilde{V}_0^{-1}\tilde{V}\tilde{O}\tilde{V}^{-1}\tilde{V}_0.
\end{eqnarray}
It reveals that the columns of $\tilde{V}_0^{-1}\tilde{V}$ are the eigenvectors of $A_2$.
The algorithm for finding natural modes is outlined as follows:
\begin{enumerate}
  \item Select an appropriate contour $C$ to enclose the nature frequencies of interest, and compute the two integrals $A_0$ and $A_1$.
  \item Obtain $\tilde{V}_0$, $\tilde{J}_0$ and $\tilde{U}_0$ through the SVD of $A_0$, then construct the matrix $A_2$.
  \item Compute the eigenvectors $\tilde{V}_0^{-1}\tilde{V}$ and eigenvalues $\tilde{O}$ from the eigenvalue problem $A_2 (\tilde{V}_0^{-1}\tilde{V})= (\tilde{V}_0^{-1}\tilde{V})\tilde{O}$.
  \item Get $\tilde{U}$ and $\tilde{V}$.
  \item Expand Eq. (\ref{CurrentMatrix}) by using $\tilde{U}$ and $\tilde{V}$:
  \begin{eqnarray}\label{pole1}
     x(\omega) \approx  \tilde{V}a(\omega) = \sum_j a_j(\omega) \tilde{V}_j,
  \end{eqnarray}
  where $\tilde{V}_j$ is a column of $\tilde{V}$,  and $a_j(\omega)$ is the expansion coefficient and the element of the column vector $a(\omega)$. $a(\omega)$ is solved by the Galerkin equation as follows,
\begin{eqnarray}\label{pole2}
\tilde{U}^{\dag}M(\omega)\tilde{V})a(\omega)=\tilde{U}^{\dag}y(\omega),\nonumber\\
     a(\omega) = (\tilde{U}^{\dag}M(\omega)\tilde{V})^{-1} (\tilde{U}^{\dag}y(\omega)).
\end{eqnarray}
\end{enumerate}
The searching of natural modes is highly dependent on the selection of the contour. The stability of the algorithm may be compromised when the contour is in proximity to branch cuts. Unlike the situation in the first approach based on SVD, as shown in Eq. (\ref{svd2}), it is important to note that $\tilde{U}$ and $\tilde{V}$ might not be orthogonal matrices in this context.

\section{Branch cuts of gold dielectric function}\label{app:branch}
When representing the dielectric function as $\epsilon = |\epsilon|e^{i\phi}$, two mathematically permissible values, $\pm\sqrt{|\epsilon|}e^{i\frac{\phi}{2}}$, arise for the refractive index $n$. To ensure the physical validity of the refractive index when considering real-valued frequencies, these is some restriction.  Specifically, it is required that $Re\left [n(\omega)\right]> 0$ and $Im\left[n(\omega)\right]>0$ to avoid divergences in the outgoing solution at infinity. It is guaranteed by the experimental refractive index of gold. However, there is some freedom in choosing the sign of the refractive index when $\omega$ is complex valued. In practice, branch cuts can be introduced to establish a continuous, single-valued branch for the refractive index $n(\omega)$. Consequently, $M(\omega)$ becomes a continuous, single-valued matrix function of $\omega$, enabling the use of various algorithms to determine the natural frequencies.  Two options for such branch cuts can be considered.  The first option is to set the negative real axis of the complex $\epsilon$ as the branch cut. In this case, we have $\epsilon = |\epsilon|e^{i\phi_1}$ with $-\pi<\phi_1\leq\pi$, resulting in $n = \sqrt{|\epsilon|}e^{i\frac{\phi_1}{2}}$. This choice ensures that the real part of the refractive index is always positive. The second option involves employing the positive real axis of $\epsilon$ as the branch cut. By introducing $\phi_2$, we express $\epsilon = |\epsilon|e^{i\phi_2}$ with $0\leq\phi_2<2\pi$ for $Re(\omega) >0$ and $2\pi\leq\phi_2<4\pi$ for $Re(\omega) <0$. Subsequently, $n = \sqrt{|\epsilon|}e^{i\frac{\phi_2}{2}}$. This choice guarantees a positive imaginary part of the refractive index for $Re(\omega) >0$. Fig.~\ref{fig:gold_dielectric} (d) and (e) illustrate the values of $\phi_1$ and $\phi_2$ on the complex plane of $\omega$, respectively. At the branch cuts, the phases exhibit discontinuities. The branch cuts connect the locations where the dielectric functions are zero and infinity, corresponding to the negative and positive real axes of $\omega$ for the two options, respectively. The values of the phases $\pi/2\pi$ from both sides of the branch cuts are clearly indicated, and the line with $0/4\pi$ is not the branch cut as the refractive index is continuous as a square root. The two options produce the identical refractive index on the real axis of $\omega$, and they can be used in calculation depending on the location of modes of interest and the path of contour integral. The second option is adopted in this work, as it shifts the branch cut away from the real axis of $\omega$.

\section{Left and right eigenvectors }\label{app:leff_right}
In this section, we demonstrate that the complex conjugate of the right eigenvector is the left eigenvector of $M$. This relationship also applies to the left and right singular vectors.

\subsection{Proof via the operator}\label{app:p1}
Define the inner product between two sets of scalar and vector potentials $\Psi^{a,b} = \left(
\begin{array}{c}
\phi^{a,b}(\mathbf{r})\\
\mathbf{A}^{a,b}(\mathbf{r})
\end{array}\right)$ as an integral performed over the closed boundary $S$,
\begin{align}
\nonumber 
<\Psi^b|\Psi^a>=\oint_{S}d{\bf s}\left[\Psi^b({\bf s})\right]^{\dag}\Psi^a({\bf s})\\
 = \oint_{S}d{\bf s}\left[\overline{\phi^b({\bf s})}\phi^a({\bf s})+\overline{\mathbf{A}^b({\bf s})}\cdot\mathbf{A}^a({\bf s})\right],
\end{align}
where $\overline{\psi^b}$ and $\overline{\mathbf{A}^b}$ are the complex conjugates of $\psi^b$ and $\mathbf{A}^b$, respectively. 

We further impose constraints on the form of $\Psi^{a,b}$ as follows:
\begin{align}
&\Psi^a = 
\begin{pmatrix}
\phi^a(\mathbf{r})=\oint_{S}ds \frac{\exp(ik_{1,2}|\mathbf{r}-\mathbf{s}|)}{|\mathbf{r}-\mathbf{s}|}\sigma_{1,2}^a(\mathbf{s}) \\
\mathbf{A}^a(\mathbf{r})=\oint_{S}d{s}\frac{\exp(ik_{1,2}|\mathbf{r}-\mathbf{s}|)}{|\mathbf{r}-\mathbf{s}|}\mathbf{h}_{1,2}^a(\mathbf{s})
\end{pmatrix}\\
&\Psi^b=\overline{\Psi^a},
\end{align}
The selection of 1 or 2 corresponds to whether $\mathbf{r}$ is located inside or outside the closed boundary $S$. It is evident that both $\Psi^{a}$ and $\Psi^{b}$ are the solutions of the Helmholtz wave equation inside and outside of the boundary,
\begin{align}
[\nabla^2 + k_{1,2}^2]\Psi^{a}(\mathbf{r})=\begin{pmatrix}0 \\\mathbf{0}\end{pmatrix},\\
[\nabla^2 + \overline{k}_{1,2}^2]\Psi^{b}(\mathbf{r})=\begin{pmatrix}0 \\\mathbf{0}\end{pmatrix}. 
\end{align}
$\sigma_{1,2}^{a,b}$ and $\mathbf{h}_{1,2}^{a,b}$ are chosen such that $\Psi^{a,b}(\mathbf{r})$ are continuous across the boundary $S$:
\begin{equation}
\Psi^{a,b}(\bf s-\varepsilon) = \Psi^{a,b}(\bf s+\varepsilon), \, s \in S \, \text{and}\, \varepsilon \to 0, 
\end{equation}
which corresponds to the source-free version of Eqs.~(\ref{foureqs_1}), and (\ref{foureqs_2}):
\begin{align}
&\phi^{a,b}(\bf s-\varepsilon) - \phi^{a,b}(\bf s+\varepsilon)=0,\label{app:phiEQ}\\
&\mathbf{A}^{a,b}(\bf s-\varepsilon) - \mathbf{A}^{a,b}(\bf s+\varepsilon)=\mathbf{0}, \label{app:AEQ}
\end{align}
where $\bf s-\varepsilon$ and $\bf s+\varepsilon$ are infinitesimally close to the boundary $S$ from inside and outside, respectively. We introduce the following notations: 
\begin{align}
    \phi_{1}^{a,b}({\bf s}) = \phi^{a,b}({\bf s-\varepsilon}), \, \phi_{2}^{a,b}({\bf s})=\phi^{a,b}({\bf s+\varepsilon}),\label{app:phi1}\\
    \mathbf{A}_{1}^{a,b}({\bf s}) = \mathbf{A}^{a,b}({\bf s-\varepsilon}), \, \mathbf{A}_{2}^{a,b}({\bf s}) =\mathbf{A}^{a,b}({\bf s+\varepsilon}).\label{app:A1}
\end{align}

From Eqs.~(\ref{foureqs_3}), and (\ref{foureqs_4}), we can define an operator $\hat{L}_{\omega}$ such that
\begin{align}\label{app:Lk}
&\hat{L}_{\omega}\Psi^a = \hat{L}_{\omega}\left(
\begin{array}{c}
\phi^a\\
\mathbf{A}^a
\end{array}\right) \\\nonumber
=&\left(
\begin{array}{c}
{\bf n}_s \cdot \nabla(\epsilon_1\phi^a_1 - \epsilon_2\phi^a_2)- ik{\bf n}_s\cdot(\epsilon_1\mathbf{A}^a_1 - \epsilon_2\mathbf{A}^a_2)\\
{\bf n}_s \cdot \nabla (\mathbf{A}^a_1-\mathbf{A}^a_2) - ik{\bf n}_s(\epsilon_1\phi^a_1 - \epsilon_2\phi^a_2)
\end{array}\right).
\end{align} 
Bear in mind that $\epsilon_{1,2}$ and $k$ are functions of $\omega$. Then we can find the relation between $\hat{L}_{\omega}$ and its adjoint operator $\hat{L}_{\omega}^{\dag}$ from the definition $<\hat{L}_{\omega}^{\dag}\Psi^b|\Psi^a>=<\Psi^b|\hat{L}_{\omega}
\Psi^a>$
\begin{widetext}
\begin{align}\label{app:Lplus}
\nonumber
&<\hat{L}_{\omega}^{\dag}\Psi^b|\Psi^a>\\
\nonumber
=&<\Psi^b|\hat{L}_{\omega}\Psi^a> = \oint_{S}d{s} \left(\begin{array}{r}\overline{\phi^b({s-\varepsilon})}\\ \overline{\mathbf{A}^b({s-\varepsilon})}\end{array}\right)^T\left(
\begin{array}{c}
{\bf n}_s \cdot \nabla(\epsilon_1\phi^a_1 - \epsilon_2\phi^a_2)- ik{\bf n}_s\cdot(\epsilon_1\mathbf{A}^a_1 - \epsilon_2\mathbf{A}^a_2)\\
{\bf n}_s \cdot \nabla (\mathbf{A}^a_1-\mathbf{A}^a_2) - ik{\bf n}_s(\epsilon_1\phi^a_1 - \epsilon_2\phi^a_2)
\end{array}\right)\\
\nonumber
=&\oint_{S}d{s}\left(
\overline{\phi^b_1}{\bf n}_s \cdot \epsilon_1\nabla\phi^a_1 -\overline{\phi^b_2}{\bf n}_s \cdot \epsilon_2\nabla \phi^a_2 + \overline{\phi^b_1}(-ik{\bf n}_s\cdot\epsilon_1\mathbf{A}^a_1) + \overline{\phi^b_2}(ik{\bf n}_s\cdot\epsilon_2\mathbf{A}^a_2)\right.\\
\nonumber
&\left.+\overline{\mathbf{A}^b_1} \cdot ({\bf n}_s \cdot \nabla) \mathbf{A}^a_1-\overline{\mathbf{A}^b_2} \cdot ({\bf n}_s \cdot \nabla) \mathbf{A}^a_2 +\overline{\mathbf{A}^b_1} \cdot(- ik{\bf n}_s\epsilon_1\phi^a_1) + \overline{\mathbf{A}^b_2} \cdot(ik{\bf n}_s\epsilon_2\phi^a_2)
\right)\\
\nonumber
=&\oint_{S}d{s}\left(
\overline{({\bf n}_s \cdot \nabla\overline{\epsilon_1}\phi^b_1)}\phi^a_1 -\overline{({\bf n}_s \cdot \nabla \overline{\epsilon_2}\phi^b_2)}\phi^a_2+\overline{i\overline{k\epsilon_1}{\bf n}_s\phi^b_1}\cdot\mathbf{A}^a_1 + \overline{-i\overline{k\epsilon_2}{\bf n}_s\phi^b_2}\cdot\mathbf{A}^a_2\right.\\
\nonumber
&\left.+\overline{(({\bf n}_s \cdot \nabla )\mathbf{A}^b_1)} \cdot \mathbf{A}^a_1- \overline{(({\bf n}_s \cdot \nabla) \mathbf{A}^b_2)}\cdot \mathbf{A}^a_2 +\overline{i\overline{k\epsilon_1}{\bf n}_s\mathbf{A}^b_1} \cdot\phi^a_1 + \overline{-i\overline{k\epsilon_1}{\bf n}_s\mathbf{A}^b_2} \cdot\phi^a_2
\right)\\
=&\oint_{S}d{s} \left(\begin{array}{c}
{\bf n}_s \cdot \nabla(\overline{\epsilon_1}\phi^b_1 - \overline{\epsilon_2}\phi^b_2)- \overline{ik}{\bf n}_s\cdot(\overline{\epsilon_1}\mathbf{A}^b_1 - \overline{\epsilon_2}\mathbf{A}^b_2)\\
{\bf n}_s \cdot \nabla (\mathbf{A}^b_1-\mathbf{A}^b_2) - \overline{ik}{\bf n}_s(\overline{\epsilon_1}\phi^b_1 - \overline{\epsilon_2}\phi^b_2)
\end{array}\right)^{\dag}
\left(
\begin{array}{r}\overline{\phi^a({s-\varepsilon})}\\ \overline{\mathbf{A}^a({s-\varepsilon})}\end{array}
\right),
\end{align}
\end{widetext}
where Eqs.~(\ref{app:phiEQ}), (\ref{app:AEQ}), (\ref{app:phi1}) and (\ref{app:A1}) are used. The divergence theorem and the integration by parts are also used, for instance,
\begin{align*}
&\oint_{S}d{s}\overline{\phi^b_1}{\bf n}_s \cdot \epsilon_1\nabla\phi^a_1 = \int_{V_1}d{\mathbf{r}}\nabla \cdot (\overline{\phi^b_1} \epsilon_1\nabla\phi^a_1)\\
=&\int_{V_1}d^3{\mathbf{r}}(\nabla \overline{\phi^b_1} \cdot \epsilon_1\nabla\phi^a_1 + \overline{\phi^b_1} \epsilon_1\nabla^2\phi^a_1)\\
=&\int_{V_1}d^3{\mathbf{r}}\left(\nabla \cdot (\epsilon_1\phi^a_1\nabla\overline{\phi^b_1})- \phi^a_1 \epsilon_1\nabla^2\overline{\phi^b_1}+ \overline{\phi^b_1} \epsilon_1\nabla^2\phi^a_1\right)\\
=&\int_{V_1}d^3{\mathbf{r}}\left(\nabla \cdot (\epsilon_1\phi^a_1\nabla\overline{\phi^b_1})+ \phi^a_1 \epsilon_1k_1^2\overline{\phi^b_1}- \overline{\phi^b_1} \epsilon_1k_1^2\phi^a_1\right)\\
=&\int_{V_1}d^3{\mathbf{r}}\nabla \cdot (\epsilon_1\phi^a_1\nabla\overline{\phi^b_1})
=\oint_{S}d{s}\overline{({\bf n}_s \cdot \nabla\overline{\epsilon_1}\phi^b_1)}\phi^a_1 .  
\end{align*}
From Eqs.~(\ref{app:Lplus}), it can be found that
\begin{align}
&\hat{L}_{\omega}^{\dag}\Psi^b = \hat{L}_{\omega}^{\dag}\left(
\begin{array}{c}
\phi^b\\
\mathbf{A}^b
\end{array}\right) \\\nonumber
=&\left(
\begin{array}{c}
{\bf n}_s \cdot \nabla(\overline{\epsilon_1}\phi^b_1 -\overline{\epsilon_2}\phi^b_2)- \overline{ik}{\bf n}_s\cdot(\overline{\epsilon_1}\mathbf{A}^b_1 - \overline{\epsilon_2}\mathbf{A}^b_2)\\
{\bf n}_s \cdot \nabla (\mathbf{A}^b_1-\mathbf{A}^b_2) - \overline{ik}{\bf n}_s(\overline{\epsilon_1}\phi^b_1 - \overline{\epsilon_2}\phi^b_2)
\end{array}\right)\\
\nonumber
=&\overline{\left(
\begin{array}{c}
{\bf n}_s \cdot \nabla(\epsilon_1\overline{\phi^b_1} -\epsilon_2\overline{\phi^b_2})- ik{\bf n}_s\cdot(\epsilon_1\overline{\mathbf{A}^b_1} - \epsilon_2\overline{\mathbf{A}^b_2)}\\
{\bf n}_s \cdot \nabla (\overline{\mathbf{A}^b_1)}-\overline{\mathbf{A}^b_2)}) - ik{\bf n}_s(\epsilon_1\overline{\phi^b_1} - \epsilon_2\overline{\phi^b_2})
\end{array}\right)}\\
\nonumber
=&\overline{\hat{L}_{\omega}\overline{\Psi^b}} = \hat{L}_{\overline{-\omega}}\Psi^b,
\end{align} 
where these facts have been used:
\begin{align*}
    \overline{\epsilon_{1,2}(\omega)} = \epsilon_{1,2}(-\overline{\omega}),\\
    \overline{ik}=\overline{i\frac{\omega}{c}}=i\frac{-\overline{\omega}}{c}=i\overline{-k}.
\end{align*}
If $\Psi^a$ is the right eigenvector, i.e., $\hat{L}_{\omega}\Psi^a = 0$, then $\Psi^b$ is the left eigenvector, i.e., $\hat{L}_{\omega}^{\dag}\Psi^b = 0$. The left eigenvector is the complex conjugate of the right eigenvector:
\begin{equation}
   \Psi^b = \overline{\Psi^a}
\end{equation}

\subsection{Proof via the matrix}\label{app:p2}
Using a similar procedure, the above relation can also be demonstrated directly from Eq.~(\ref{Mdefiniion}) or the left side of Eq.~(\ref{current_equations}) where, upon discretization, all terms can be treated as matrices.  Suppose $G_2\mathbf{h}_2$ is the right eigenvector with the eigenvalue $\lambda(\omega)$ which can potentially be zero at the natural resonance $\omega^*$:
\begin{align}\nonumber
&M(\omega)G_2\mathbf{h}_2\\
\nonumber
=&\left(\Delta + k^2(L_1-L_2){\bf n }_s \Gamma ^{-1} {\bf n }_s\cdot(L_1-L_2)\right)G_2\mathbf{h}_2\\
=&\lambda(\omega)G_2\mathbf{h}_2.
\end{align}
First we demonstrate that $\Delta$ is a symmetric matrix:
\begin{align*}
\nonumber
&\Delta = H_1G_1^{-1}-H_2G_2^{-1} \\
\nonumber
=& (G_1^{-1})^TG_1^TH_1G_1^{-1}-(G_2^{-1})^TG_2^TH_2G_2^{-1}\\
=& (G_1^{-1})^TH_1^TG_1G_1^{-1}-(G_2^{-1})^TH_2^TG_2G_2^{-1}\\
=& (G_1^{-1})^TH_1^T-(G_2^{-1})^TH_2^T=\Delta^T,
\end{align*}
where $H_{1,2}^TG_{1,2} = G_{1,2}^TH_{1,2}$ is used. It could be derived as follow; note that matrix multiplication here is interpreted as the integral over the closed surface,
\begin{align*}
\nonumber
&\left(G_{1,2}^TH_{1,2}\right)_{ij} = (G_{1,2})_{li}(H_{1,2})_{lj}\\
=&G_{1,2}(s_l, s_i)ds_i(\mathbf{n}_{s_l} \cdot \nabla)G_{1,2}(s_l, s_j)ds_j \frac{1}{\delta s_l}\delta s_l\\
=&\frac{1}{\delta s}\oint_{S} ds_l G_{1,2}(s_l, s_i)ds_i(\mathbf{n}_{s_l} \cdot \nabla)G_{1,2}(s_l, s_j)ds_j\\
=&\frac{1}{\delta s}\int_{V_{1,2}}d^3{\mathbf{r}}\nabla \cdot \left[G_{1,2}(\mathbf{r}, s_i)ds_i\nabla G_{1,2}(\mathbf{r}, s_j)ds_j\right]\\
=&\frac{1}{\delta s}\int_{V_{1,2}}d^3{\mathbf{r}}\left\{\nabla \cdot \left[\nabla G_{1,2}(\mathbf{r}, s_i)ds_i G_{1,2}(\mathbf{r}, s_j)ds_j\right]\right.\\
&\left.-\nabla^2 G_{1,2}(\mathbf{r}, s_i)ds_i G_{1,2}(\mathbf{r}, s_j)ds_j \right.\\
&\left. + G_{1,2}(\mathbf{r}, s_i)ds_i \nabla^2 G_{1,2}(\mathbf{r}, s_j)ds_j\right\}\\
=&\frac{1}{\delta s}\int_{V_{1,2}}d^3{\mathbf{r}}\nabla \cdot \left[\nabla G_{1,2}(\mathbf{r}, s_i)ds_i G_{1,2}(\mathbf{r}, s_j)ds_j\right]\\
=&\frac{1}{\delta s}\oint_{S}ds_l\left[\mathbf{n}_{s_l} \cdot \nabla G_{1,2}(s_l, s_i)ds_i\right]  G_{1,2}(s_l, s_j)ds_j\\
=&(H_{1,2})_{li}(G_{1,2})_{lj}=\left(H_{1,2}^TG_{1,2}\right)_{ij},
\end{align*}
where it is assumed that the area of each surface mesh element is $\delta s$.
Similarly, $\Gamma$ is also a symmetric matrix. Then it can be found that $\overline{G_2\mathbf{h}_2}$ is the left eigenvector as below
\begin{align}\nonumber
&\left(\overline{G_2\mathbf{h}_2}\right)^{\dag}M(\omega)\\
\nonumber
=&\mathbf{h}_2^TG_2^T\left(\Delta +  k^2(L_1-L_2)\cdot{\bf n }_s \Gamma ^{-1} {\bf n }_s(L_1-L_2)\right)\\
\nonumber
=&\mathbf{h}_2^TG_2^T\left(\Delta^T + k^2(L_1-L_2)\cdot {\bf n }_s (\Gamma ^{-1})^T {\bf n }_s(L_1-L_2)\right)\\
\nonumber
=&\left(M(\omega)G_2\mathbf{h}_2\right)^T\\
\nonumber
=&\left(\lambda(\omega)G_2\mathbf{h}_2\right)^T\\
=&\left(\overline{G_2\mathbf{h}_2}\right)^{\dag}\lambda(\omega)
\end{align}
Hence, the left eigenvector is the complex conjugate of the right eigenvector for the matrix $M(\omega)$. In the context of singular value decomposition, it can be demonstrated as well that the left and right singular vectors form a complex conjugate pair.

\section{Mie theory}\label{app:mie}
For a sphere in a uniform background, vector spherical harmonics serve as the normal mode solutions of Helmholtz equation without external source in spherical coordinates $(r, \theta, \phi)$ with the spherical unit vectors $\hat{e}_r$, $\hat{e}_{\theta}$ and $\hat{e}_{\phi}$. These vector spherical harmonics can be defined as follows \cite{bohren_absorption_1998}
\begin{align}
{\bf M}_{emn} = &\frac{-m}{\sin\theta}\sin m\phi P^m_n(\cos\theta)z_n(\rho)\hat{e}_{\theta}\nonumber\\
                &-\cos m\phi\frac{dP^m_n(\cos\theta)}{d\theta}z_n(\rho)\hat{e}_{\phi},\\
{\bf M}_{omn} = &\frac{m}{\sin\theta}\cos m\phi P^m_n(\cos\theta)z_n(\rho)\hat{e}_{\theta}\nonumber\\
                &-\sin m\phi\frac{dP^m_n(\cos\theta)}{d\theta}z_n(\rho)\hat{e}_{\phi},\\
{\bf N}_{emn} = &\frac{z_n(\rho)}{\rho}\cos m\phi\, n(n+1) P^m_n(\cos\theta)\hat{e}_{r}\nonumber\\
                &+\cos m\phi \frac{dP^m_n(\cos\theta)}{d\theta}\frac{1}{\rho}\frac{d}{d\rho}\left[ \rho z_n(\rho)\right]\hat{e}_{\theta}\nonumber\\
                &-m\sin m\phi\frac{P^m_n(\cos\theta)}{\sin\theta}\frac{1}{\rho}\frac{d}{d\rho}\left[ \rho z_n(\rho)\right]\hat{e}_{\phi},\\
{\bf N}_{emn} = &\frac{z_n(\rho)}{\rho}\sin m\phi\, n(n+1) P^m_n(\cos\theta)\hat{e}_{r}\nonumber\\
                &+\sin m\phi \frac{dP^m_n(\cos\theta)}{d\theta}\frac{1}{\rho}\frac{d}{d\rho}\left[ \rho z_n(\rho)\right]\hat{e}_{\theta}\nonumber\\
                &+m\cos m\phi\frac{P^m_n(\cos\theta)}{\sin\theta}\frac{1}{\rho}\frac{d}{d\rho}\left[ \rho z_n(\rho)\right]\hat{e}_{\phi},
\end{align}
where $\rho = k_jr$, and the subscripts $o$ and $e$ denote odd and even modes, respectively. $m(0\leq l \leq n)$ and $n$ are the mode order numbers, originating from the associated Legendre polynomials $P_n^m(\cos\theta)$.  $z_n$ is the spherical Bessel functions. For each $n$, there are $2n+1$ modes ($n+1$ odd modes and $n$ even modes). There exists a relationship 
\begin{align}
{\bf N} = \frac{\nabla \times {\bf M}}{k\sqrt{\epsilon_j\mu_j}},\; {\bf M} = \frac{\nabla \times {\bf N}}{k\sqrt{\epsilon_j\mu_j}}.
\end{align}
 
The field of incident plane wave, the scattering field and the field within the sphere, including electric field, magnetic field and vector potential,  can all be expanded into a series of these vector spherical harmonics. ${\bf M}_{omn}$ and ${\bf M}_{emn}$ do not have a radial part. Here, we use TE to denote the mode when it is satisfied that ${\bf E\propto M}_{o(e)mn}$, ${\bf H\propto N}_{o(e)mn}$, and ${\bf A\propto M}_{o(e)mn}$; TM denotes the mode when it is satisfied that ${\bf E\propto N}_{o(e)mn}$, ${\bf H\propto M}_{o(e)mn}$, and ${\bf A\propto N}_{o(e)mn}$. The total scattering cross section is the sum of the contributions from all these modes \cite{bohren_absorption_1998},
\begin{equation}
    \sigma_s = \frac{2\pi}{k^2\epsilon\mu}\sum_{n=1}^{\infty}(2n+1)(|a_n|^2+|b_n|^2),
\end{equation}
where $a_n$ and $b_n$ are the coefficients for TM and TE modes, respectively \cite{bohren_absorption_1998}. 

These vector spherical harmonics are functions of $kr$. The angular part of ${\bf M}_{o(e)mn}$ is independent of $kr$, and we have 
\begin{align}
{\bf M}_{o(e)mn}(k_1r) = \frac{z_n(k_1r)}{z_n(k_2r)}{\bf M}_{o(e)mn}(k_2r).
\end{align}
For ${\bf N}_{o(e)mn}$, we can have  
\begin{align}
{\bf N}_{o(e)mn}(k_1r) = \alpha {\bf N}_{o(e)mn}(k_2r) + \beta {\bf N}_{o(e)mn}(k_3r),
\end{align}
where $\alpha$ and $\beta$ are the coefficients. Therefore, for a TE mode, the electric field or the vector potential (${\bf\propto M}_{o(e)mn}$) at any wavenumber $k_1$ can be expressed in terms of the electric field or the vector potential at a fixed wavenumber $k_2$. For a TM mode, the electric field or the vector potential (${\bf\propto M}_{o(e)mn}$) at any wavenumber has to be expressed as a combination of the electric field or the vector potential at any two fixed wavenumbers. This explains that six SVD modes are required to express the TM 1 modes.
\bibliography{apssamp}
\end{document}